\documentclass[lettersize,journal]{IEEEtran}
\usepackage{amsmath,amsfonts}
\usepackage{graphicx}
\usepackage{cite}
\usepackage{booktabs}
\usepackage{arydshln}
\usepackage{url}
\usepackage{textcomp}
\usepackage{eso-pic}

\usepackage{xcolor}

\usepackage{algorithm}
\usepackage{algpseudocode}

\begin{document}

\AddToShipoutPictureFG*{%
  \AtPageLowerLeft{%
    \put(\LenToUnit{0.5\paperwidth},\LenToUnit{12pt}){%
      \makebox[0pt][c]{\parbox[b]{\textwidth}{\centering\fontsize{6}{7.2}\selectfont
        \copyright~2026 IEEE. Personal use of this material is permitted. Permission from IEEE must be obtained for all other uses, in any current or future media, including reprinting/republishing this material for advertising or promotional purposes, creating new collective works, for resale or redistribution to servers or lists, or reuse of any copyrighted component of this work in other works. This is the accepted version of the article published in IEEE Transactions on Visualization and Computer Graphics, doi: \pdfstartlink attr{/Border[0 0 0]} user{/Subtype/Link/A<</S/URI/URI(https://doi.org/10.1109/TVCG.2026.3750350)>>}10.1109/TVCG.2026.3750350\pdfendlink.}}}}}

\title{Physics-Driven Extraction of Dynamic Fluid Asset from Monocular Video with Generative 3DGS}

\author{Zhiwei~Zhao, Yin~Zhao, Minchen~Li, and~Yixin~Hu%
\thanks{Z. Zhao and Y. Zhao are with Tencent, Shenzhen, China (e-mail: \mbox{zhaozw483@gmail.com}; \mbox{alantzhao@tencent.com}).}%
\thanks{M. Li is with Carnegie Mellon University, Pittsburgh, PA, USA (e-mail: \mbox{minchernl@gmail.com}).}%
\thanks{Y. Hu is with Tencent America, New York, NY, USA (e-mail: \mbox{yixinhu.yh@gmail.com}). (Corresponding author: Yixin Hu.)}}

\markboth{IEEE Transactions on Visualization and Computer Graphics}%
{Zhao \MakeLowercase{\textit{et al.}}: Physics-Driven Extraction of Dynamic Fluid Asset from Monocular Video with Generative 3DGS}

\maketitle

\begin{abstract}
Existing methods for dynamic 3D scene synthesis from video rely on per-frame deformation fields optimized through pixel-level reconstruction losses. While effective for closed systems such as elastic solids, these approaches face challenges with open systems like fluids where material continuously enters and exits the scene, topology changes freely, and no stable point correspondence exists across frames.
We propose a simulation-driven alternative that optimizes the parameters of a fluid simulator to reproduce dynamics observed in a single-view video, rather than fitting per-point deformation to pixel values. Our framework leverages generative 3D Gaussian Splatting (3DGS) models to obtain initial fluid geometry including unseen surfaces, estimates a physically grounded velocity field from surface motions with a physics-constraint correction algorithm, and converts it into a divergence-free, grid-based representation. A differentiable Affine Particle-In-Cell (APIC) simulation then optimizes effective simulation parameters, such as viscosity, boundary conditions, and inflow/outflow velocities, to match the observed dynamics.
Because the dynamics are governed by forward simulation, the output assets are inherently editable, interactive, and temporally extendable,
which reconstruction-based approaches cannot readily provide. 
Stage-wise ablations confirm that each stage is necessary, and hyperparameter sweeps show the pipeline to be robust 
without sensitive per-scene tuning. Our method handles various liquid fluids, including inviscid and viscous types, demonstrating the potential for automatically creating simulation-ready 3D fluid assets from readily available internet videos.

\end{abstract}

\begin{IEEEkeywords}
Fluid simulation, 3D Gaussian Splatting, physics-based simulation, physical parameter optimization, dynamic fluid assets.
\end{IEEEkeywords}

\section{Introduction}

\IEEEPARstart{C}{ombining} 3D generation with editing, animation, and simulation is reshaping digital content creation. Coupling generation with physics-based dynamics is hardest for fluids: most existing work targets rigid bodies and viscoelastic materials, whose outcomes are more stable and predictable, yet demand for fluid assets remains high, since creating them by hand is labor-intensive given the intricacy of fluid dynamics and the precision required of simulation parameters.

\begin{figure}[!t]
  \centering
  \includegraphics[width=0.48\textwidth,clip,trim=0pt 0pt 1450pt 0pt]{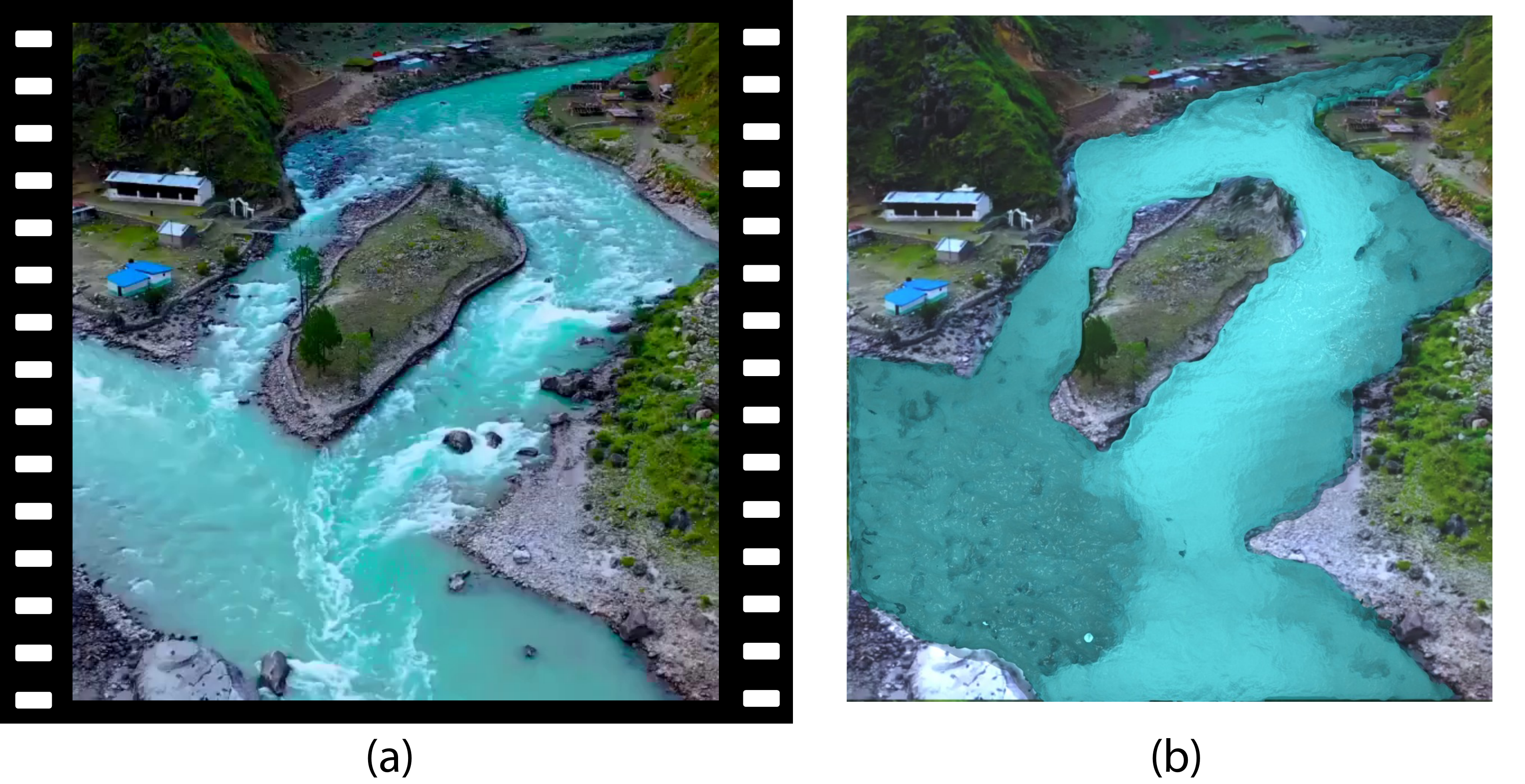}\\[0pt]
  \hspace{6pt}
  \includegraphics[width=0.44\textwidth,clip,trim=1600pt 0pt 0pt 0pt]{figs/teaser_new_compressed.pdf}

  \centering
  \caption{Our method takes a single-view video of fluid as input as shown in (a) and automatically extracts a high-quality ready-to-use dynamic fluid asset shown in (b). Users can extend the geometry and add interacting objects, as demonstrated in the edited result in (c), which is then rendered in (d) with original extracted appearance.}
  \label{fig:teaser}
\end{figure}

A natural question arises: can we extract simulation-ready 3D fluid assets directly from casually captured videos?
Recent advances in dynamic 3D reconstruction offer a tempting starting point, but these methods are generally designed for a different class of problems. Dynamic Gaussian Splatting methods learn per-frame deformation fields by minimizing pixel-level reconstruction losses. This paradigm works well for closed deformable systems such as elastic bodies where each material point persists throughout the sequence and per-vertex tracking is meaningful. However, fluids are \textit{open systems}: material continuously enters and exits the scene through inflow and outflow, topology changes freely, and no stable point correspondence exists across frames.
To apply pixel-level reconstruction to fluids, one would need to track material that does not persist in the scene. And even if per-pixel fitting succeeds for the training camera, the resulting model may not generalize. The deformation field explains the observed view but is not constrained to produce physically correct 3D motion from other angles. Downstream capabilities crucial for asset creation (e.g., changing viewing directions, adding interactions, and extending motion duration) will benefit from the underlying dynamics being physically grounded, beyond visual plausibility alone.

Meanwhile, recent work has embedded physics simulators into reconstructed Gaussians for forward animation, and has recovered fluid dynamics from visual observation through differentiable rendering and learned dynamics models. Section~\ref{sec:related_work} discusses these and their relation to our work.

Content creators seek fluid assets that are not only visually plausible but also editable, interactive, and integrable into existing 3D scenes.
Based on these considerations and insights,
we formulate the problem as \textit{physical parameter optimization}. Rather than fitting a deformation field to minimize photometric error, we seek the simulation parameters of a physics-based fluid solver that, when executed as a forward simulation, reproduce the dynamics observed in the input video. The output is not a fixed per-frame deformation but a resimulatable physical model, together with an initial appearance estimation, that naturally supports novel-view rendering, interaction, temporal extension, and intuitive parameter editing.
Concretely, we approximate the fluid geometry with 3D Gaussians (3DGS) obtained through generative methods and ensure geometry consistency across frames. The volumetric velocity field is then estimated from the generated 3D geometry and the motion present in the video, leveraging optical flow corrected by a physics-derived constraint.
Building on the APIC method~\cite{APIC}, a hybrid Eulerian-Lagrangian scheme that conserves both linear and angular momentum, we design a differentiable framework to optimize the simulation parameters, including viscosity, boundary conditions, and inflow/outflow velocities, to best recover the dynamics from the video.
Our experiments show that the proposed method produces high-quality results for fluids with convectional flow motion. Comparisons demonstrate the necessity and efficacy of the physics-based optimization: reconstructing precise geometry but manually specifying simulation parameters produces notably less accurate results that diverge over time, suggesting that inferring the dynamics from observation is both necessary and beneficial.

This prioritizes physical dynamics over pixel-level appearance, a deliberate choice rooted in the nature of the problem. Per-pixel reconstruction can achieve high image similarity, but it is pseudo-physical and guarantees neither generalization to novel views nor downstream editability. For viscous fluids, which flow steadily and change topology little, tracking remains feasible, and our method recovers an appearance closely resembling the video. For inviscid fluids with high boundary flux, such as flowing rivers, material enters and leaves too rapidly for strict tracking. Since surface texture at any moment is a consequence of the underlying velocity field, recovering the dynamics is more informative than fitting the appearance.
The recovered velocity field and simulation parameters yield spatially and temporally coherent 3D fluids matching the video dynamics, exposed through intuitive quantities (viscosity, gravity, boundary coefficients) that are straightforward to tune. The extracted assets are compatible with standard graphics engines such as Blender and Unreal Engine, enabling integration into game development and virtual production pipelines.

The main contributions can be summarized as follows:

\begin{itemize}
    \item
    We propose a novel framework that formulates 3D fluid asset extraction from single-view videos as a physical parameter optimization problem, providing simulation-ready outputs with inherent editability, interactivity, and temporal extendability.
    \item We design fluid-customized shape and motion estimation strategies to produce coherent geometry and physically grounded velocity fields, including a mainstream correction algorithm with 2D-divergence constraints derived from the incompressibility condition.
    \item
    We realize a differentiable grid velocity evolution procedure building upon the APIC method to optimize simulation parameters, compatible with general grid-based or hybrid simulation methods.
\end{itemize}

\section{Related Work}
\label{sec:related_work}

\subsection{3D Generation with Gaussian Splatting}
3D Gaussian Splatting (3DGS)~\cite{kerbl3Dgaussians} adopts a point-based radiance field, using 3D Gaussian primitives to represent scenes. It has emerged as a prevalent research topic in 3D representation~\cite{zhang2024rade,huang20242d, Yu2024GOF,lu2024scaffold,yu2024mip} due to its ability to depict high-quality geometry and textures in novel view synthesis.

3DGS provides a new perspective not only for real-time scene reconstruction but also for 3D generation. DreamGaussian~\cite{tang2023dreamgaussian} optimizes a 3DGS through score distillation using a pre-trained text-to-image diffusion model.
Recent methods~\cite{he2025gvgen, zhang2024gaussiancube, zhou2024diffgs} investigate directly training diffusion models on Gaussian splats for higher efficiency. However, the direct methods may struggle to handle real-world inputs, like fluid objects, since they are usually trained on synthetic 3D datasets.
LGM~\cite{tang2025lgm} transforms the single-view generation problem into a multi-view 3DGS reconstruction task using a pre-trained single-view to multi-view 2D diffusion model.
The most recent transformer-based methods~\cite{zou2024triplane, xiang2024structured} are proposed to achieve faster generation with higher quality. TriplaneGaussian~\cite{zou2024triplane} creates a point cloud from a single image and uses hybrid triplane-Gaussian representation to greatly accelerate the generation.
TRELLIS~\cite{xiang2024structured} employs rectified flow transformers unifying structured latent to get high-quality results.
We use such model as an upstream component, taking the initial 3D geometry with appearance from it and embedding physical dynamics through our simulation-based pipeline.

\subsection{Dynamic Scene Reconstruction with Deformation Fields}
\label{sec:related_deformation}
The neural radiance expression of objects with dynamics has long been studied for NeRF~\cite{mildenhall2021nerf} systems, including deformation capture by canonical and displacement fields~\cite{park2021nerfies, park2021hypernerf} and dynamic 3D synthesis~\cite{ne1gao2021dynamic,ne2li2021neural,ne3liu2022devrf,ne4pumarola2021d,ne5qiao2023dynamic}.
Similarly, for the later proposed 3D Gaussian Splatting, dynamics properties have been imposed on the explicit radiance representation of Gaussians.
Leveraging the high fitting ability of Gaussian particles, many studies have explored dynamic scene reconstruction guided by image-based losses from video~\cite{ren2024l4gm, wu20244d, yan20244d, gao2024gaussianflow, novotny2022keytr, duan20244d}.
These methods learn per-frame deformation fields, typically parameterized by MLPs or decomposed neural encodings, that warp a canonical Gaussian representation to match each observed frame, optimized through photometric reconstruction losses.
This paradigm has achieved impressive results for general dynamic scenes, particularly for objects that maintain stable topology and persistent material identity across time (e.g., deforming elastic bodies, articulated objects, or faces).
However, applying these methods to fluid dynamics poses fundamental challenges.

In fluids, material identity is transient and topology changes freely through splashing, merging, and fragmentation~\cite{bridson2015fluid, foster2001practical}, so the per-Gaussian assumption of a persistent identity does not hold. The same difficulty is acknowledged for NeRF-based deformation fields~\cite{park2021hypernerf}.
Moreover, deformation fields optimized solely through photometric losses may satisfy the visual objective without respecting incompressibility or momentum conservation~\cite{chu2022physics, li2023nvfi}.
They therefore suffice for view interpolation within the observed sequence, but do not guarantee physically consistent dynamics from novel viewpoints, and extending the motion or introducing interactions remains difficult.

These observations reflect a difference in problem setting rather than a shortcoming of these methods, which target general dynamic scene reconstruction with faithful visual reproduction as the objective. Ours is a complementary setting whose downstream goal, an editable and simulation-ready fluid asset, places different demands on the representation.

\subsection{Physics-Based Simulation and Its Integration with 3D Gaussians}
In physics-based simulation, fluids are governed by the Navier-Stokes equations and often simulated using Eulerian grids and/or Lagrangian particles~\cite{bridson2015fluid},
with different considerations of viscosity~\cite{bardos2007euler,doering1995applied}. Numerical discretization schemes have also advanced these computations a lot~\cite{foster1996realistic, tome1994gensmac}.
For simulation stability, Stam~\cite{stablefluid99} introduces the concept of semi-Lagrangian advection, which brings up the idea of hybrid methods.
Hybrid methods~\cite{PIC,FLIP, APIC, PolyPIC,jiang2016material, mls_mpm} combine
advantages of Lagrangian and Eulerian schemes, representing fluid by particles while computing dynamics on grids.
Among these, the Affine Particle-In-Cell (APIC) method~\cite{APIC} is notable for its conservation of both linear and angular momentum through the affine velocity transfer, which we adopt in our framework for its suitability to fluid dynamics.

Recent works have begun integrating physics-based simulation with 3D Gaussian representations, demonstrating the value of combining physical correctness with the rendering efficiency of 3DGS.
PhysGaussian~\cite{xie2023physgaussian} pioneered this direction by integrating the Material Point Method (MPM) framework with reconstructed 3DGS, enabling versatile dynamic behaviors for elastic, viscoplastic, and granular materials.
GaussianSplashing~\cite{feng2025splashing} combines Position-Based Dynamics (PBD) with advanced Gaussian rendering, achieving real-time fluid-solid interaction synthesis with natural fluid appearances.
PhysMotion~\cite{tan2024physmotion} proposes a framework for animating 3DGS generated from a single-view image using differentiable MPM, with sequential rendering results composited through an inversion and diffusion process.
RainyGS~\cite{dai2025rainygs} and FieryGS~\cite{shen2026fierygs} extend the approach to rain and fire in captured scenes.
These works demonstrate the effectiveness of embedding physics into Gaussian representations for producing physically consistent, interactive animations. These works operate in a \textit{forward simulation} setting where the physical parameters (material properties, forces, boundary conditions) are specified by the user. While this offers full creative control, for an artist who wants to reproduce the specific dynamics observed in a reference video, manually tuning the high-dimensional parameter space through trial and error can be challenging.

A complementary direction seeks to \textit{infer} physical parameters or dynamics from visual observation.
Simplicits~\cite{modi2024simplicits} uses a neural field of learnable weights for reduced-order simulation.
PhysDreamer~\cite{zhang2024physdreamer} optimizes material parameters through a differentiable MPM simulation to reproduce dynamic Gaussians that behave similarly to input videos, providing an inspiring example of physics-based inverse optimization, though its formulation is designed for elastic solids rather than fluids.
BAGS~\cite{peng2024bags} learns the weights of imposed Gaussian Ellipsoid Neural Bounds to animate reconstructed objects according to input videos.
Hong et al.~\cite{hong2026physics} treat each Gaussian as a material point and fit time-varying constitutive parameters under optical-flow supervision.
In our work, we build upon the idea of physics-based inverse optimization and extend it to the fluid domain. We embed generated Gaussians in physics compatible with the APIC method and design a differentiable pipeline to optimize the simulation parameters to best recover the observed fluid dynamics from video.

\subsection{Fluid Dynamics Recovery from Videos}
\label{sec:related_velocity}
Recovering fluid velocity fields from visual observation is a long-standing challenge.
Optical flow~\cite{opticalflow2003} can effectively capture pixel-level motion and has been utilized in neural reconstructions~\cite{li2021neural,du2021neural} for general dynamic scenes. Targeting specific fluid textures can noticeably improve learning of dynamics and re-simulation accuracy~\cite{guan2022neurofluid,liu2023inferring}.
Generally, multi-view inputs are required for 3D velocity reconstruction, and novel-view fluid synthesis typically relies on at least sparse viewpoints for neural networks to infer the unseen sides~\cite{multiokabe2015fluid,multieckert2018coupled,multieckert2019scalarflow}. Consequently, previous work has seldom focused on reconstructing the complete velocity field from a single view.

Several works have explored incorporating physics priors into fluid dynamics recovery.
DVP~\cite{deng2023learning} successfully encodes vortex features to learn the specific eddy dynamics of fluid, leveraging the divergence-free property inherent in vortex representations, though it is restricted to 2D profiles.
Other works~\cite{franz2021global, chu2022physics, yu2024inferring} extend to 3D space with strict physical constraints and volume rendering, achieving high-fidelity reconstruction of certain fluid types.
NeuroFluid~\cite{guan2022neurofluid} jointly learns a particle transition model and a particle-driven neural renderer from multi-view video, pioneering the unsupervised learning of particle-based fluid dynamics from visual observation.
Liu et al.~\cite{liu2023inferring} integrate a differentiable Euler simulator with volumetric rendering for inverse fluid inference, using a learned pressure solver to enable end-to-end optimization.
More recently, FluidNexus~\cite{gao2025fluidnexus} addresses 3D fluid reconstruction and prediction from a single video. It employs a novel-view video synthesizer to generate multi-view supervision and optimizes a particle representation using Position-Based Fluids (PBF) constraints within an image-driven optimization loop. This approach achieves compelling results for gaseous flows (e.g., smoke) and demonstrates interactive capabilities, representing a thoughtful integration of physical priors with visual reconstruction.

\begin{figure*}[t]
  \centering
  \includegraphics[width=\textwidth]{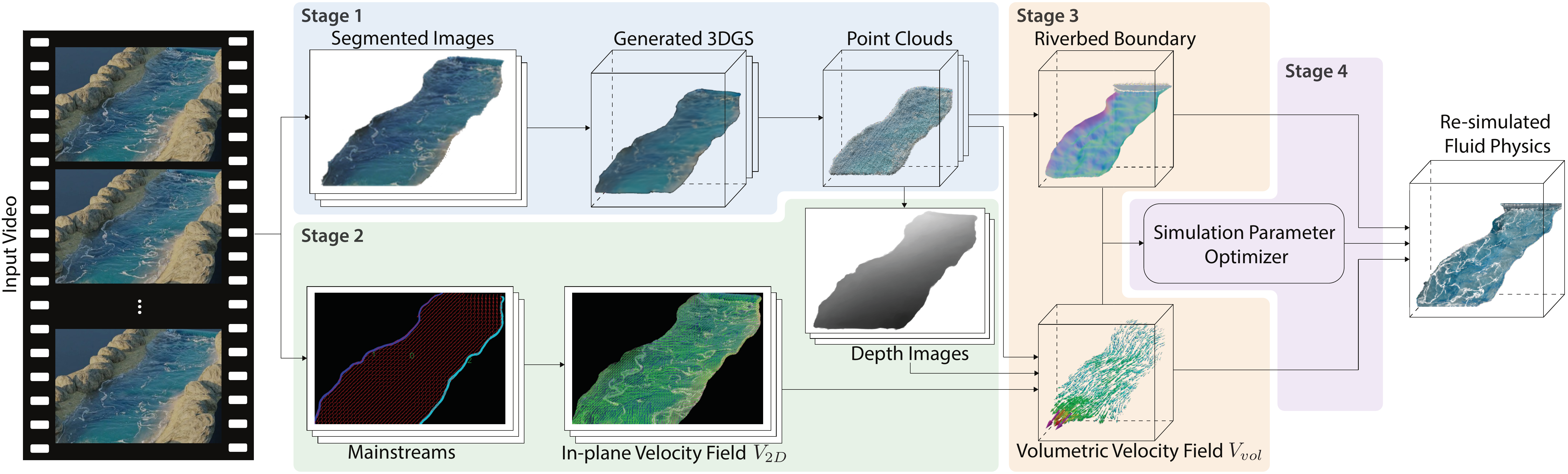}  
  \caption{Pipeline overview. Stage~1: generating and preprocessing 3DGS from input frames. Stage~2: estimating surface motion vectors via optical flow with mainstream correction and 2D-divergence projection. Stage~3: constructing the volumetric velocity field with boundary-layer estimation and constrained projection. Stage~4: optimizing simulation parameters through differentiable APIC evolution. The output is then post-processed for rendering.}
  \label{fig:overview}
\end{figure*}

Our work shares the high-level goal of recovering 3D fluid dynamics from a single-view video, but differs significantly in the target fluid type and the role of the simulation.
We focus on liquids (both inviscid and viscous). Unlike gaseous flows, liquids exhibit free-surface dynamics, boundary-layer effects, and terrain interactions.
Rather than using a physics simulator as a soft constraint within an image-driven optimization, we treat forward APIC simulation as the primary driver and extract its parameters directly from the video. Consequently, our output is inherently a complete simulation rather than a fitted deformation field.
Our approach uses generative 3DGS models for geometry estimation and solves for surface motions to initialize velocity.
This offers a different trade-off compared to multi-view video diffusion, which typically incurs higher inference costs in memory and time and is harder to control.
We view these as complementary approaches, and address the gap between them by estimating the physical parameters of a liquid simulator from a single-view video.

\section{Method}

\subsection{Overview}
Our method takes a single-view video of fluid as input and outputs a simulation-ready 3D asset, consisting of optimized geometry, initial appearance, physical parameters, and boundary conditions that can be re-simulated, edited, and temporally extended.
The pipeline (Fig.~\ref{fig:overview}) comprises four stages:

\noindent
\textbf{1: 3D Fluid Geometry Estimation} (Sec.~\ref{sec:geometry}).
We employ image-to-3DGS generative models to obtain an initial 3D Gaussian representation from the input frame. The generated Gaussians are denoised, volumetrically densified, and unified across selected frames to ensure geometric consistency. A single-view appearance optimization then refines resolution and fidelity on the settled point cloud.

\noindent
\textbf{2: Surface Motion Vector Estimation} (Sec.~\ref{sec:surface-motion}).
We estimate 2D screen-space motion from inter-frame optical flow, addressing its inherent limitations (e.g., the failure in homogeneous-texture regions) through a mainstream correction strategy and a projective physical constraint derived from 3D incompressibility. The corrected 2D motion vectors are then inversely mapped onto 3D surfaces via the camera.

\noindent
\textbf{3: Volumetric Velocity Field Construction} (Sec.~\ref{sec:volumetric}).
With surface motions providing the outer boundary condition, we estimate the near-wall velocity profile using boundary-layer functions from computational fluid dynamics. The complete volumetric velocity field is then solved via a constrained projection method that enforces divergence-free flow throughout the domain.

\noindent
\textbf{4: Differentiable  Parameter Optimization} (Sec.~\ref{sec:optimization}).
We transfer the velocity field to a grid-based representation compatible with the APIC scheme. A differentiable grid velocity evolution, with a PeRCNN-based pressure solver for efficient backpropagation, enables gradient-based optimization of the simulation parameters to match the observed dynamics. 

Algorithm~\ref{alg:pipeline} overviews the pipeline end to end. Stages
connect through the particle set $\mathcal{P}$, the screen-space field
$\mathbf{u}_{2D}$ and volumetric field $\mathbf{V}_{\text{vol}}$, so each
can be re-run on its own.

\subsection{Background}
\subsubsection{Generative 3D Gaussian Splatting}
Thanks to its differentiable volume rendering, 3DGS~\cite{kerbl3Dgaussians} can be employed in image-to-3D generation with textures aligned to the conditioning image. The generation procedure $\mathcal{G}$ maps a single image to a set of Gaussian primitives:
$\mathcal{G}:\{C\}_{w,h} \rightarrow {\{\mathbf{x},\sigma,\mathbf{A},\mathbf{f}\}}_p$,
where $\{C\}_{w,h}$ represents pixels of the conditioned image at position $(w, h)$.
$\mathbf{x}$, $\sigma$, $\mathbf{A}$, and $\mathbf{f}$ denote the position, opacity, covariance matrix, and spherical harmonic features of each Gaussian particle $p$, respectively. The input image can be recovered through discretized splatting rendering:
$
    \sum_{i \in \mathcal{N}}  \alpha_i \boldsymbol{SH}(r|_{w,h};\mathbf{f}_i)                      \prod_{j=1}^{i-1} (1-\alpha_j )\rightarrow \{C\}_{w,h},
$
where $\alpha_i$ is the product of $\sigma_i$ and the projected 2D Gaussian density at the intersection of kernel $i$ with the ray $r|_{w,h}$, and $\boldsymbol{SH}$ denotes the color computed from spherical harmonic features $\mathbf{f}_i$. This formulation is equivalent to a depth-and-opacity weighted average of particle colors along each ray.
Recent single-image-to-3DGS methods~\cite{zou2024triplane, xiang2024structured, tang2025lgm, szymanowicz2024splatter, xu2024instantmesh} run in seconds, and the point-based nature of 3DGS is directly compatible with particle-based fluid simulation, making them suitable upstream components.

\begin{algorithm}[!t]
\caption{Fluid asset extraction from a single-view video}
\label{alg:pipeline}
\small
\begin{algorithmic}[1]
\Require frames $\{C^t\}$, reference frame index $t_0$
\Ensure simulation parameters $\Theta$, particles $\mathcal{P}$

\Statex \textbf{Stage 1: 3D fluid geometry} \hfill(Sec.~\ref{sec:geometry})
  \For{each selected frame $t$}
    \State $\{\mathbf{x},\sigma,\mathbf{A},\mathbf{f}\}^t \gets \mathcal{G}(C^t)$
      \Comment{generation}
    \State discard particles with $\sigma<\tau_\sigma$ \Comment{denoising}
    \State add $\mathbf{x}$ at ray-cast interior cells
      \Comment{volumetric filling}
  \EndFor
  \State $\mathcal{P}\gets\bigcup_t\{\mathbf{x}\}^t$ \Comment{cross-frame union}
  \State refine $\{\sigma,\mathbf{A},\mathbf{f}\}$ of $\mathcal{P}$ against $C^{t_0}$
    \Comment{appearance optimization}

\Statex \textbf{Stage 2: surface motion} \hfill(Sec.~\ref{sec:surface-motion})
  \State $\mathbf{d}\gets\textsc{OpticalFlow}(C^{t},C^{t+\Delta t_{\text{video}}})$
  \State $M\gets\{k:|\mathbf{d}_k|\ge\tau_u\}$ \Comment{reliable measurements}
  \ForAll{$k\notin M$}
    \State $\mathbf{d}_k\gets\sum_{i\in B(k)}w(i)
           \max\!\big(0,\vec{n}_k\!\cdot\!\hat{\mathbf{d}}_i\big)^{p}\mathbf{d}_i$
      \Comment{mainstream}
  \EndFor
  \State relax $\mathbf{u}_{2D}\!\gets\!\mathbf{d}$ towards
         Eq.~\ref{eq::div2DV} \Comment{2D constraint}
  \State $\mathbf{u}_{3D}\gets(\mathbf{u}_{2D},\Delta z)$, with $\Delta z$ the depth
         change along $\mathbf{u}_{2D}$
  \State $\mathbf{V}_{\text{surf}}\gets
         \big(\mathbf{S}\circ(\mathbf{P}\mathbf{W})+\mathbf{T}\big)^{-1}
         \circ\mathbf{u}_{3D}$ \Comment{inverse mapping to 3D}

\Statex \textbf{Stage 3: volumetric velocity} \hfill(Sec.~\ref{sec:volumetric})
  \State detect inlet, outlet and terrain cells \Comment{B.C. regions}
  \State $\vec{v}_i\gets$ propagate $\mathbf{V}_{\text{surf}}$ inward, radius $R$
  \State scale $|\mathbf{v}_{\text{base}}|/V_{\infty}$ within $y<\delta$
         by Eq.~\ref{eq::wallFn} \Comment{wall functions}
  \State $\mathbf{V}_{\text{vol}}\gets\vec{v}_i+\Delta\mathbf{v}_j$ s.t.
         $C_i\!:\!(\nabla\!\cdot\!\mathbf{v})_i\!=\!0$,
         Eq.~\ref{eq::divProj} \Comment{projection}

\Statex \textbf{Stage 4: parameter optimization} \hfill(Sec.~\ref{sec:optimization})
  \State $\Theta\gets\arg\min_\Theta\mathcal{L}
         \big(\textsc{DiffSim}(\mathbf{V}_{\text{vol}};\Theta),
         \mathbf{V}_{\text{vol}}\big)$, Eq.~\ref{eq::optLoss}
\end{algorithmic}
\end{algorithm}

\subsubsection{Affine Particle-In-Cell (APIC) Method}
APIC~\cite{APIC} is a hybrid Eulerian-Lagrangian simulation technique that augments each particle with a local affine velocity field. Unlike standard PIC~\cite{PIC}, which suffers from excessive numerical dissipation, or FLIP~\cite{FLIP} which introduces noise, APIC exactly conserves both linear and angular momentum during particle-grid transfers, making it well suited for fluid dynamics. At time step~$n$, particle velocities are transferred to the grid together with their affine components:

\begin{equation}
\begin{aligned}
    (m\vec{v})_i^{n} &= \sum_p  w_{ip} m_p [\mathbf{v}^n_p + \mathbf{C}_p^n (\mathbf{x}_i - \mathbf{x}_p^n) ] \\
\vec{v}_i^{n} &= (m\vec{v})_i^{n} \big/ \sum_p w_{ip} m_p ,
\end{aligned}
\label{eq::particle_map}
\end{equation}
where $i$ and $p$ denote grid and particle coordinates respectively. $m$ is the mass, $\vec{v}_i$ is the grid velocity and $\mathbf{v}_p $ is the particle velocity. $w$ is the weight function, such as quadratic B-Spline kernel, and $\mathbf{C}$ is the affine matrix associated with each particle.
Assuming the simulated media is incompressible, the velocity change $\frac{\partial \vec{v}}{\partial t}$ can be computed on the grid by the following equation:
\begin{equation}
\begin{aligned}
\frac{\partial \vec{v}}{\partial t}
=&-\frac{\nabla p}{\rho} +\nabla\cdot(2\nu\textbf{S}_{ij})+g \\
\textbf{S}_{ij}=&\frac{1}{2}[\nabla \vec{v}+(\nabla \vec{v})^\mathsf{T}],
\end{aligned}
\label{eq::v}
\end{equation}
where $\nabla p$ is the gradient of pressure, $\rho$ is the density, $\nabla \cdot$ is the divergence operator, $g$ is the acceleration of body forces like gravity. $\nu$ denotes kinematic viscosity, modeling the viscous stress by the strain rate $\textbf{S}_{i,j}$ which is the deviatoric part of the velocity gradient.
A splitting scheme~\cite{chorin1967numerical} decouples this PDE: the body force is applied explicitly, the velocity is then relaxed to be divergence-free through the Poisson pressure equation, and the viscous stress is finally computed from the strain rate~\cite{bridson2015fluid}.
The updated grid velocity is then transferred back to the particles as states of the next time step $n+1$, and the corresponding affine matrix is simultaneously updated:
\begin{equation}
\begin{aligned}
    \mathbf{v}_p^{n+1} &= \sum_i w_{ip}{\vec{v}_i}^{n+1} =\sum_i w_{ip} \big(\vec{v}_i^{n}+ \big( \frac{\partial \vec{v}_i}{\partial t} \big)^n \Delta t \big),\\
     \mathbf{C}_p^{n+1} &= \frac{4}{\Delta x^2} \sum_i w_{ip} {\vec{v}_i}^{n+1} (\mathbf{x}_i - \mathbf{x}_p^n)^T ~(\text{quadratic kernel}).
\end{aligned}
\end{equation}
Finally, the convective effect of simulated media is achieved on the particles by
$
    \mathbf{x}_p^{n+1} = \mathbf{x}_p^n + \mathbf{v}_p^{n+1} \Delta t
$.

\subsection{3D Fluid Geometry Estimation}
\label{sec:geometry}

Since the input consists of a single-view video, we rely on generative image-to-3DGS models~\cite{zou2024triplane, xiang2024structured, tang2025lgm} to obtain 3D geometry. The choice of generative model is modular. Our pipeline is compatible with different generators and benefits from their continued improvements. We use TriplaneGS~\cite{zou2024triplane} by default as it best retains the geometry and appearance from inputs like scenic clips, and also demonstrate results with TRELLIS~\cite{xiang2024structured}.

However, generative models are typically trained on surface-dominated 3D datasets, producing sparse Gaussians concentrated on visible surfaces. For fluid simulation, we require dense, volumetrically filled geometry with consistent structure across frames. We first finalize the geometry through denoising, densification, and cross-frame union, then optimize the appearance on the resulting point cloud so that both geometry and appearance are settled before motion estimation begins.

\subsubsection{Denoising and Volumetric Densification}
\begin{figure}[!t]
\centering
\begin{minipage}{0.4\linewidth}
  \centering
  \includegraphics[width=\linewidth]{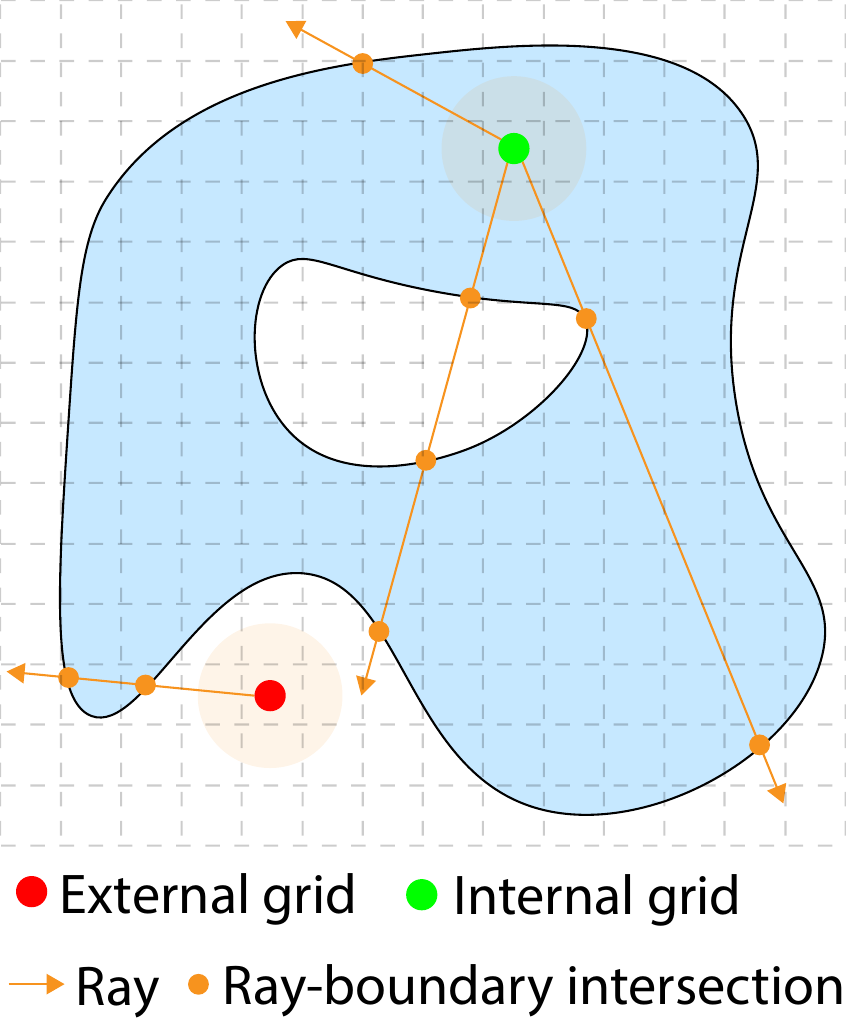}
\end{minipage}
\begin{minipage}{0.42\linewidth}
  \caption{
  Volumetric filling. For each sampled position in a cell, multiple rays are cast in random directions. A point is classified as interior if the majority of rays yield an odd intersection count, avoiding misclassification from individual rays that tangentially graze  inner vacancies.
  }
  \label{fig:fill}
\end{minipage}
\end{figure}

The generated Gaussians are first denoised by pruning primitives with low opacity or heavily stretched covariance.
As shown in Fig.~\ref{fig:fill}, we then locate the 3DGS bounding volume into a regular grid and insert a Gaussian at each interior cell by Jittered sampling. The interior check is determined by an odd-intersection ray-casting test against the outer hull of the existing Gaussians.

\subsubsection{Cross-Frame Geometric Union}
\label{sec::unionn}
Generative models do not guarantee geometric consistency across frames: adjacent frames may produce substantially different backside geometries despite similar foreground conditioning. For flowing liquids on fixed terrain, this inconsistency impairs the construction of physically grounded velocity fields.
We address this by performing a union operation on the generated Gaussians within a frame window of size $N$,
forming a geometrically consistent batch for subsequent steps.
This operation primarily processes the backside geometry (riverbed, terrain beneath the fluid) and the silhouette edges (e.g., river banks), where generative predictions vary most across frames. At the silhouette, some frames may exhibit splashing or overflow beyond the bank. The union retains these particles, but they carry no velocity in other frames and thus do not affect downstream simulation. The camera-facing fluid surface is not altered by the union, preserving natural surface variation.
The window size $N$ is determined dynamically based on inter-frame motion intensity, which can be estimated from optic flows.

\begin{figure}[!t]
    \centering
    \includegraphics[width=\linewidth]{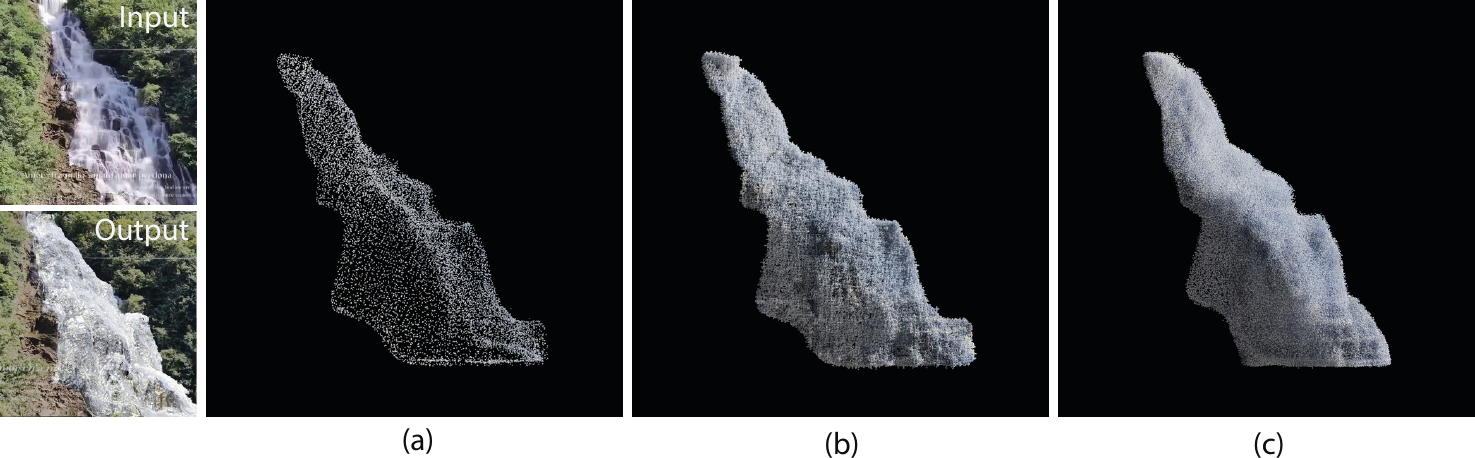}
    \caption{
    The filling operation inserts Gaussians into sparse generated 3DGS in (a) and outputs a dense 3DGS in (b). The union strategy (c) references backside and silhouette-edge geometry across few frames, producing more temporally consistent terrain and further densifies the interior.}
    \label{fig:denoisefill}
\end{figure}

The effects of densification and union are illustrated in Fig.~\ref{fig:denoisefill}.  Densification fills the interior vacancy to prevent nonphysical collapse during simulation, while the union consolidates backside and edge geometry across frames.

\subsubsection{Appearance Optimization}
\label{sec: postopt}
\begin{figure}[!t]
    \centering
    \includegraphics[height = 3cm, width=0.4\linewidth]{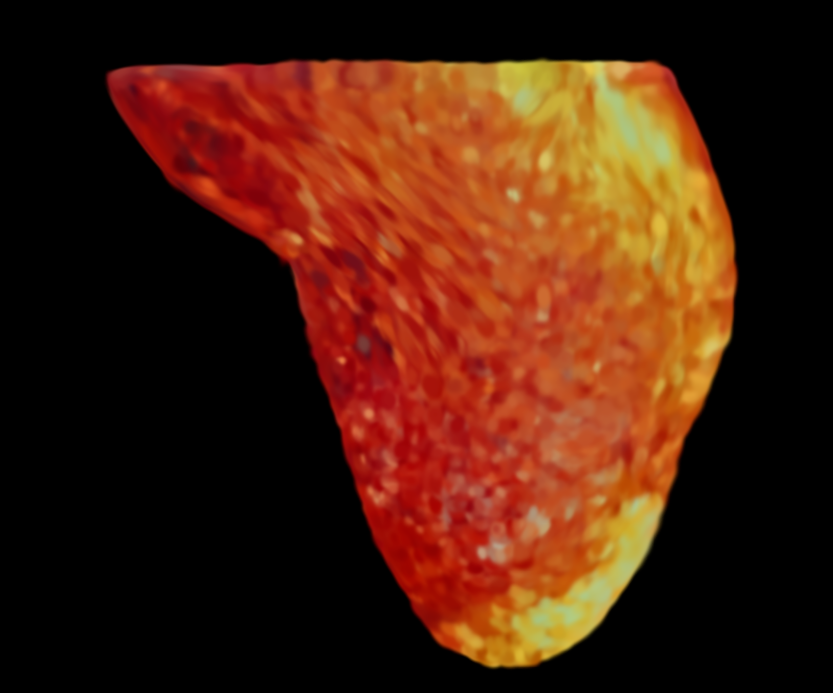}
    \includegraphics[height = 3cm,width=0.4\linewidth]{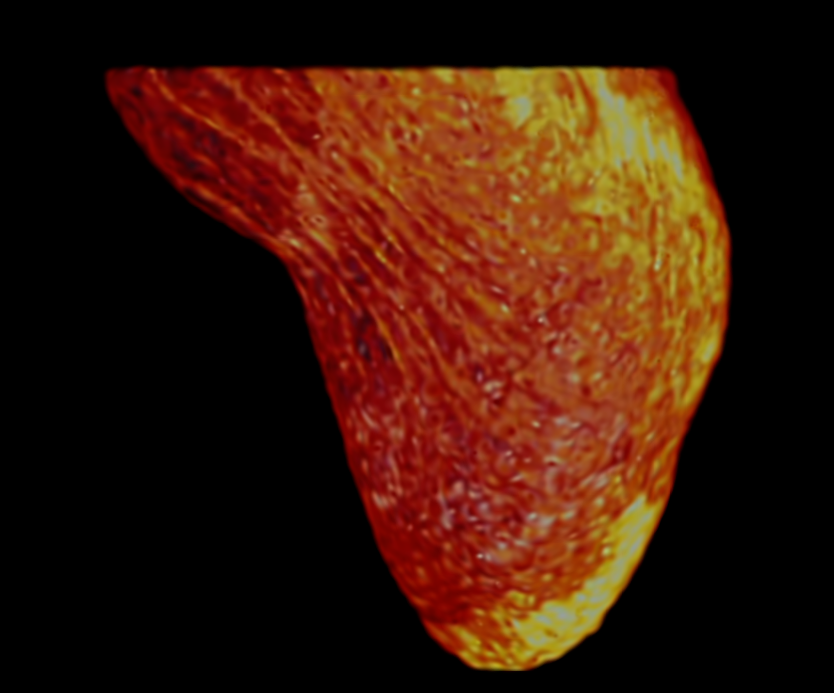}
    \includegraphics[width=0.4\linewidth]{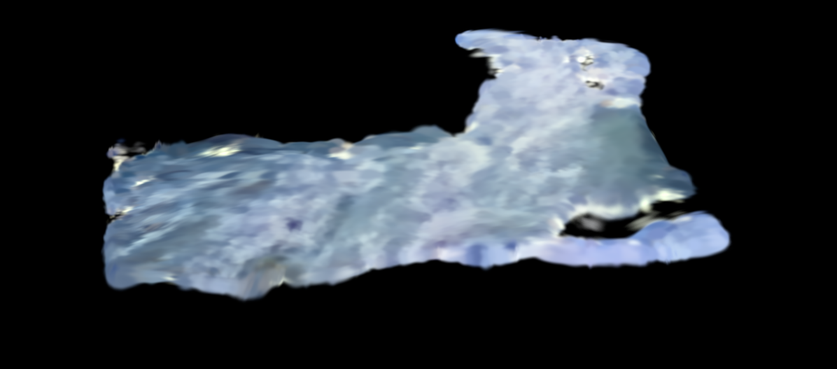}
    \includegraphics[width=0.4\linewidth]{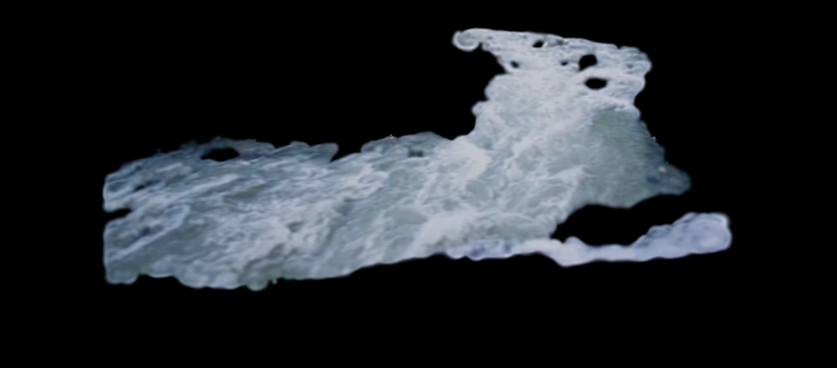}
    \caption{Preprocessed results for generated 3DGS. The left column is the direct generation, and the right column is processed with the single-view optimization.}
    \label{fig:gsopt}
\end{figure}

With the geometry finalized through the above steps, we perform a fast single-view optimization using frames from the input video to refine the appearance and resolution of the Gaussians (Fig.~\ref{fig:gsopt}). This ensures that the point cloud carries accurate per-pixel color information before surface motion estimation.
Using only one viewpoint reduces computation time but necessitates a carefully designed loss function to mitigate overfitting:
\begin{equation}
    \mathcal{L} = \lambda_{1}\mathcal{L}_{img} + \lambda_{2}\mathcal{L}_{aniso} + \lambda_{3}\mathcal{L}_{vol} + \lambda_{4}\mathcal{L}_{scl} + \lambda_{5}\mathcal{L}_{lumi}.
    \label{eq::loss_optGS}
\end{equation}
$\mathcal{L}_{img}$ denotes the image loss adopted from original 3D Gaussian Splatting~\cite{kerbl3Dgaussians}. $\mathcal{L}_{aniso}$ (anisotropic loss) and $\mathcal{L}_{vol}$ (volume loss) are implemented the same as Gaussian Splashing~\cite{feng2025splashing}. $\mathcal{L}_{scl}$ represents the geometric mean of Gaussian ellipsoids' three scale components, encouraging finer particle details. $\mathcal{L}_{lumi}$ is the luminance loss, penalizing oversaturated colors in harmonic features.
During optimization, we prune Gaussians whose colors closely match the background, effectively eliminating misrepresented rocks or reefs near the water surface. A lower densification threshold than the original 3DGS is used to encourage finer detail.
\subsection{Surface Motion Vector Estimation}
\label{sec:surface-motion}

With the 3D geometry established, we next estimate the motion of the fluid surface from the input video. We adopt the term \textit{motion vector} rather than velocity for this stage, as the estimated quantities are inter-frame displacements in screen space that are later converted to physical velocities.

\subsubsection{Screen-Space Motion Detection}
\label{sec:optflow-limitations}

We use optical flow~\cite{opticalflow2003} to extract 2D screen-space displacements, denoted as $\mathbf{u}_{2D}$, inferring material displacements in Normalized Device Coordinates (NDC) from  pixel motion between consecutive frames. 
The $z$-depth information from the preprocessed 3DGS provides the out-of-plane component.

Optical flow, however, has well-known limitations. Its constraint equation gives only one scalar equation per pixel for a vector displacement, a fundamental indeterminacy known as the \textit{aperture problem}~\cite{opticalflow2003}: only the component along the local image gradient (the \textit{normal flow}) is reliably measured, while the tangential component remains undetermined. The result stays visually plausible but need not coincide with the true 3D material motion, and may violate the underlying physics that downstream stages rely on.

A more severe issue on fluid surfaces is that homogeneous or slowly varying texture produces near-zero image gradients and thus no detectable flow. Reliable detection concentrates on waves, splashing, and texture boundaries, which reveal the bulk motion only partially.

\subsubsection{Mainstream-Guided Interpolation}
\label{sec:mainstream}
\begin{figure}[!t]
    \centering
    \includegraphics[width=0.95\linewidth]{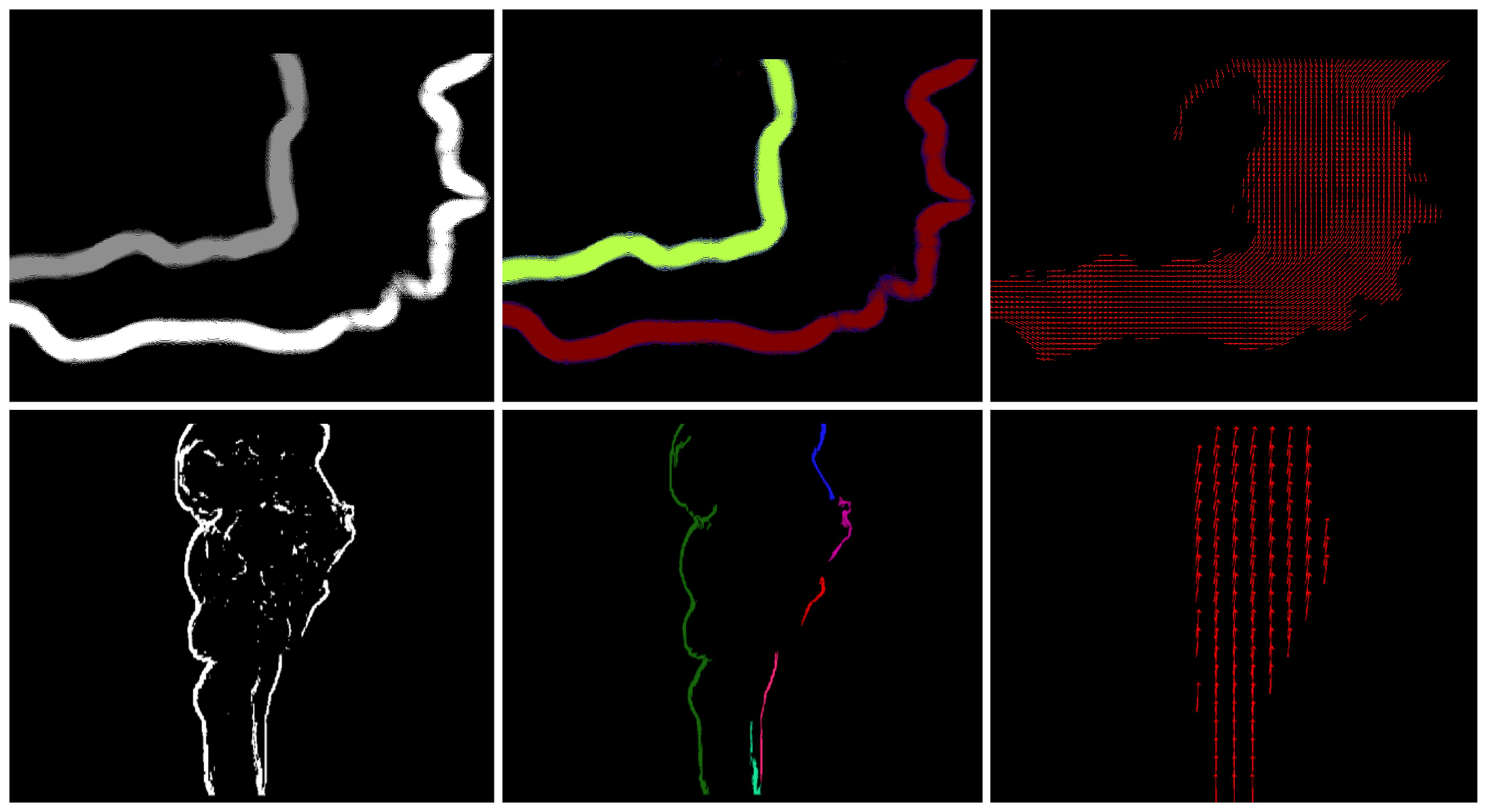}
    \caption{
Mainstream direction estimation. Edge detection identifies the fluid region boundary, connected contours are traced 
to group co-direction flows. 
Boundary flows are propagated inward, producing a smooth mainstream field.    }
    \label{fig: mainstream}
\end{figure}

Where detection fails, the underlying bulk flow is typically smooth and characterized by a dominant direction, which motivates our \textit{mainstream correction}. We take the mainstream direction $\vec{n}$ from the scene geometry and the motion magnitude from neighboring regions where optical flow succeeds.
For terrain-bounded flows (e.g., rivers), we apply edge detection to the segmented fluid region, trace connected boundary contours, and propagate the flow directions, yielding a smooth mainstream aligned with the channel geometry.
For free-surface flows not confined by terrain (e.g., poured fluids), we overlap consecutive frames to extract the principal direction from the moving average.
The interpolation then fills the failed regions as:
\begin{equation}
    \mathbf{d}_{k}=\sum_{i \in B(k)}w(i) \max\!\bigg(0,\;\vec{n}_k\cdot\frac{\mathbf{d}_{i}}{|\mathbf{d}_{i}|} \bigg)^p             \cdot \mathbf{d}_{i},
\end{equation}
where $\mathbf{d}$ denotes the motion vector in NDC space (derived from the optical flow) 
, $i$ indexes pixels with valid optical flow, and $k$ indexes pixels where optical flow fails. $B(k)$ defines the spatial neighborhood of $k$ with distance-based weights $w(i)$. The dot product between $\vec{n}_k$ and the normalized $\mathbf{d}_i$ selects neighbors whose motion aligns with the mainstream, suppressing contributions from local splashing or cross-flow disturbances. The power $p$ controls the suppressing strength. 
This is applied only where optical flow fails, preserving detected turbulence in regions with valid measurements.
Figure~\ref{fig: mainstream} shows the process for calculating the mainstream directions.

\subsubsection{2D-Divergence Constraint}
\label{sec:div-free-2d}
Interpolation alone is insufficient for high-fluidity surface.
As aforementioned, the aperture problem makes it hard to determine flow tangential to image gradient. Together with measurement noise, physical consistency is not guaranteed.
We derive a constraint that the 2D screen-space motion must satisfy, based on the 3D incompressibility (divergence-free) condition.

Under perspective projection, a 3D point $(x, y, z)$ in camera space maps to screen coordinates $\mathbf{x}_{2D}\!=\!(s,t)$:
$
s \!=\! x/z$ and $t \!=\! y/z.
$
Differentiating in time yields the screen-space velocity $\mathbf{u}_{2D} = (u_s, u_t)$:
\[
u_s = \frac{v_x - s\,v_z}{z}, \qquad u_t = \frac{v_y - t\,v_z}{z}.
\]
To compute the 2D divergence, we relate screen-space to world-space derivatives via 
$\partial x/\partial s =z$ and $\partial y/\partial t =z$. Each component is expanded as:
\[
\frac{\partial u_s}{\partial s} = z\,\frac{\partial}{\partial x}\!\left(\frac{v_x - s\,v_z}{z}\right), \qquad
\frac{\partial u_t}{\partial t} = z\,\frac{\partial}{\partial y}\!\left(\frac{v_y - t\,v_z}{z}\right).
\]
Summing both terms:
\begin{equation}
\begin{aligned}
\nabla_{2D} \cdot \mathbf{u}_{2D} =& \frac{\partial v_x}{\partial x} + \frac{\partial v_y}{\partial y} - \frac{2v_z}{z} \\
&- \frac{s}{z}\,\frac{\partial v_z}{\partial s} - \frac{t}{z}\,\frac{\partial v_z}{\partial t} -\frac{u_s}{z}\frac{\partial z}{\partial s} - \frac{u_t}{z}\frac{\partial z}{\partial t}.
\end{aligned}
\label{eq:expanded}
\end{equation}
Substituting the 3D incompressibility condition
\[
\frac{\partial v_x}{\partial x} + \frac{\partial v_y}{\partial y} = -\frac{\partial v_z}{\partial z}
\]
and dropping the $\frac{\partial v_z}{\partial z}$ term, which is considered negligible near a free surface in quasi-steady flow, we obtain (full derivation in supplementary section):
\begin{equation}
    \nabla_{2D} \cdot \mathbf{u}_{2D} = 
-\frac{1}{z} \langle \mathbf{x}_{2D}, \nabla_{2D}  \dot z \rangle 
- \frac{2}{z}\dot z
-\frac{1}{z} \langle \mathbf{u}_{2D}, \nabla_{2D}  z      \rangle
,
\label{eq::div2DV}
\end{equation}
where $z$ is depth, and $\dot z$ is the out-of-plane velocity component. This equation depends solely on inter-frame depth information from the 3DGS. In regions where the surface is approximately fronto-parallel to the camera, depth varies little and the right-hand side approaches zero, recovering approximate 2D incompressibility.

\begin{figure}[!t]
    \centering
    \includegraphics[width=\linewidth]{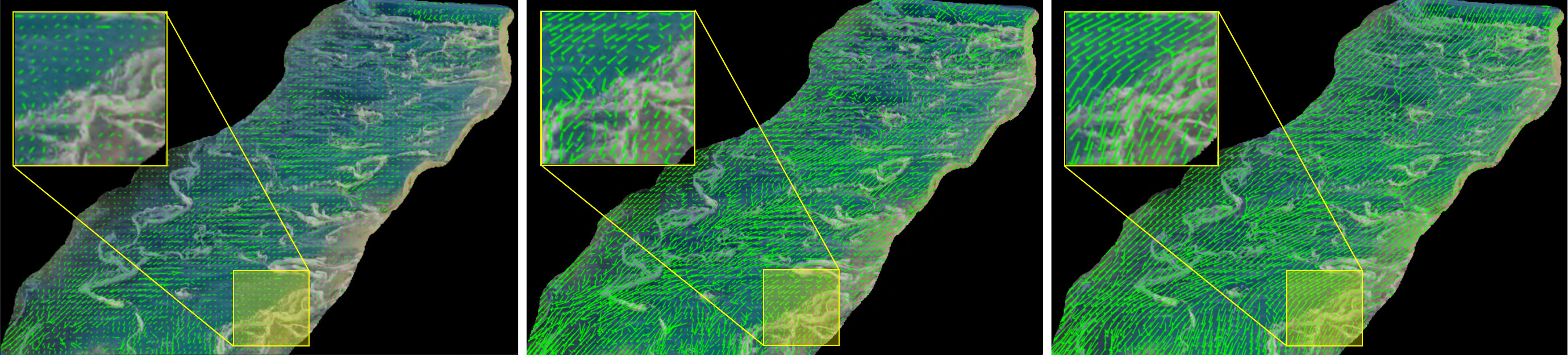}
    \caption{The unguided result shows vanished velocity in regions such as the one highlighted in the close-up (left). After applying mainstream-guided neighboring interpolation, we obtain the result shown in the middle. With further physics-constrained velocity correction, we achieve a more consistent velocity field (right).}
    \label{fig: opflow}
\end{figure}

Since the optical flow data in the failure regions does not form a well-posed boundary, the constraint alone cannot uniquely determine $\mathbf{u}_{2D}$. We therefore employ an iterative solver: $\mathbf{u}_{2D}$ is initialized using the mainstream interpolation (Sec.~\ref{sec:mainstream}) and then iteratively relaxed to satisfy Eq.~\ref{eq::div2DV}. 
Figure~\ref{fig: opflow} demonstrates that this two-step strategy, interpolation followed by divergence constraint, recovers physically plausible motion where raw optical flow detects near-zero displacement, while interpolation alone yields noisier dynamics.

\subsubsection{Inverse Mapping to 3D Surface}
\label{sec:inverse-map}
The corrected 2D motion vectors are first expanded with the depth change obtained by back-tracing along the motion itself, 
forming a 3D displacement in NDC space. 
This 3D displacement is then inversely mapped to world coordinates via the camera projection:
$
    \mathbf{v}_{\text{surf}} = \big(\mathbf{S} \circ (\mathbf{P} \mathbf{W})+\mathbf{T} \big)^{-1}\circ \mathbf{u}_{3D}
$,
where $\mathbf{u}_{3D} = (\mathbf{u}_{2D},\, \Delta z)$ is the concatenated 3D NDC displacement, $\mathbf{P}$ is the view projection matrix, $\mathbf{W}$ is the world-to-camera matrix, and $\mathbf{T}$, $\mathbf{S}$ denote translation and scaling from any preprocessing applied to input images before 3DGS generation.
This yields a physically grounded  surface motion field $\mathbf{V}_{\text{surf}}$ for subsequent volumetric estimation.

\subsection{Volumetric Velocity Field Construction}
\label{sec:volumetric}

The surface motion field $\mathbf{V}_{\text{surf}}$ from Stage~2 provides boundary conditions on the visible surface, but physical simulation requires a complete volumetric velocity field $\mathbf{V}_{\text{vol}}$ throughout the fluid domain. Constructing $\mathbf{V}_{\text{vol}}$ involves two steps: estimating the velocity near solid boundaries (the riverbed or container walls) and solving for a divergence-free interior field.

\subsubsection{Boundary Velocity via Wall Functions}
\label{sec:wall-functions}
The enclosure surface velocity $\mathbf{V}_{\text{encl}} = \mathbf{V}_{\text{surf}} \cup \mathbf{V}_{\text{base}}$ must be fully determined before computing $\mathbf{V}_{\text{vol}}$.
While the no-slip condition requires zero velocity at solid boundaries analytically, the discrete simulation grid allows a non-zero velocity at the cells nearest the boundary.
To estimate this near-wall velocity, we employ wall functions from computational fluid dynamics, which model the velocity profile within the boundary layer based on grid resolution and fluid type.

\begin{figure}[!t]
    \centering
    \includegraphics[width=0.45\linewidth]{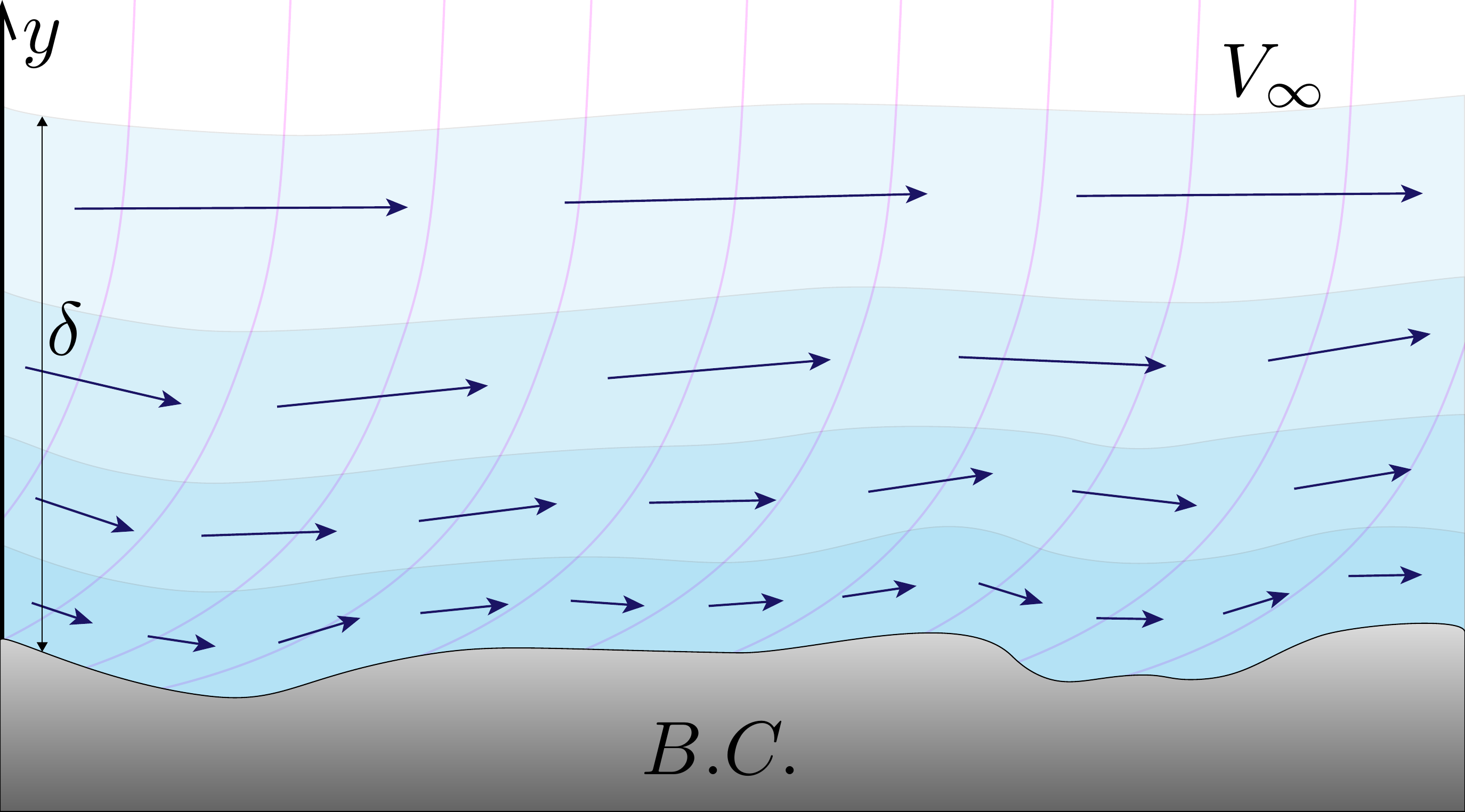}
    \includegraphics[width=0.4\linewidth]{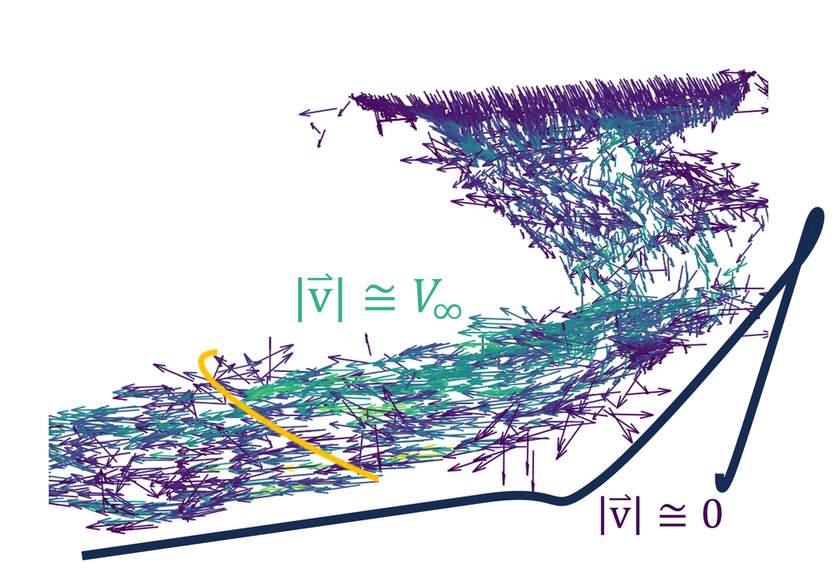}
    \caption{Boundary-layer velocity profile. Left: schematic illustration showing velocity dampening near the riverbed. Right: a simulated volumetric velocity field from our test scene, where darker color indicates smaller magnitude.}
    \label{fig: bc_layer}
\end{figure}

Specifically, we adopt the third-order polynomial approximation of the laminar boundary-layer profile:
\begin{equation}
    \frac{|\mathbf{v}_{\text{base}}|}{V_{\infty}} = \frac{3}{2} \left( \frac{y}{\delta} \right) - \frac{1}{2} \left( \frac{y}{\delta} \right)^3,
\label{eq::wallFn}%
\end{equation}
where $V_{\infty}$ is the magnitude of the corresponding surface velocity, $y$ is the distance from the wall, and $\delta$ is the boundary-layer thickness. The thickness $\delta$ is set dynamically by fluid type: for flowing rivers, a larger $\delta$ captures the gradual velocity gradient across the riverbed; for free-surface fluids, $\delta$ is limited to sub-cell scale, introducing minimal damping. The direction of $\mathbf{v}_{\text{base}}$ follows the mainstream rather than the local surface velocity.
This ensures more realistic velocity transition near solid boundaries (Fig.~\ref{fig: bc_layer}), avoiding the introduction of nonphysical supervision that could interfere with downstream optimization.

\subsubsection{Divergence-Free Constrained Projection}
\label{sec:constrained-proj}
With the complete enclosure velocity determined, we solve for a divergence-free 3D velocity field throughout the interior using a constrained projection method. The boundary-layer estimates from the previous step provide better-posed  boundary conditions.

For each grid cell $i$, we impose the divergence-free constraint $C_i: (\nabla \cdot \mathbf{v})_i = 0$ and iteratively update the velocity:
\begin{equation}
\begin{aligned}
      &C_i:~ (\nabla \cdot \mathbf{v})_i=0, ~~~~~~~~~ \mathbf{v}_j^{n+1} = \mathbf{v}_j^{n} + \Delta \mathbf{v}_j,\\
      &\Delta \mathbf{v}_j = \sum_i \lambda_i\nabla_{v_j}C_i, ~~~~~~~ \lambda_i = -\frac{C_i(\mathbf{v}_1,\dots,\mathbf{v}_n)}{\sum_k|\nabla_{v_k}C_i|^2+\epsilon}.
\end{aligned}
\label{eq::divProj}
\end{equation}
Each constraint $C_i$ depends on the velocities $\mathbf{v}_k$ of neighboring cells.
The velocity correction $\Delta \mathbf{v}_j$ is accumulated from the gradients of all constraints that $\mathbf{v}_j$ influences, and the relaxation parameter $\epsilon$ ensures stable convergence. Unlike the Chorin projection method~\cite{chorin1967numerical}, this formulation directly projects velocities onto the divergence-free manifold without forming an intermediate pressure field.

\begin{figure}[!t]
    \centering
    \includegraphics[width=\linewidth]{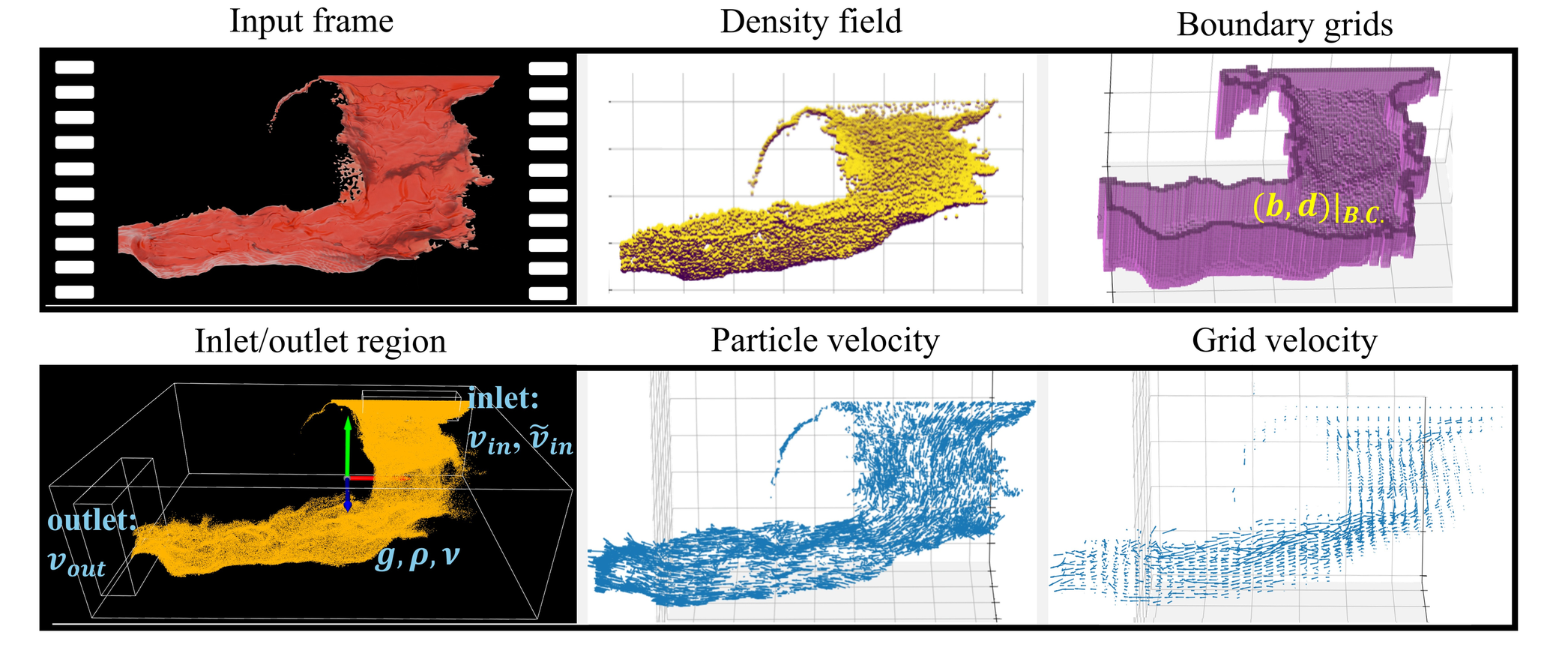}
    \caption{
    From the preprocessed 3DGS, we calculate the density field and detect boundaries in the grid. The optimization process then initializes velocities at  inlet/outlet regions, and transfer velocities from particles to grid.
    }
    \label{fig: grid_bc}
\end{figure}

\subsection{Differentiable  Parameter Optimization}
\label{sec:optimization}

The volumetric velocity field from previous stage provides a physically grounded initialization but does not yet determine the simulation parameters needed for forward simulation. In this stage, we optimize these parameters by differentiating through a grid-based APIC velocity evolution.

\subsubsection{Grid-Based Velocity Evolution}\label{sec:grid-evolution}
We transfer the particle velocity to a regular grid using Eq.~\ref{eq::particle_map} with the affine matrix $\mathbf{C}$ initialized to zero (Fig.~\ref{fig: grid_bc} bottom). This grid-based representation offers two advantages: (1)~it enables efficient tracking of fluid dynamics between frames via fixed grid coordinates, and (2)~solving the fluid equations (Eq.~\ref{eq::v}) alongside grid-based convection makes the dynamics differentiable with respect to the simulation parameters (see supplementary section for detailed gradient derivations).

\subsubsection{PeRCNN-Based Pressure Solver}
\label{sec:percnn}
The Poisson pressure equation must be solved at each simulation step to enforce incompressibility. 
Traditional iterative solvers (e.g., Jacobi, conjugate gradient) are not natively differentiable. Backpropagating through the full iteration unrolls a deep computation graph leading to memory consumption and possible gradient degradation.
To enable gradient-based optimization, we adopt a modified PeRCNN~\cite{rao2021embedding} architecture as a differentiable substitute for the pressure solver.

\begin{figure}[!t]
    \centering
    \includegraphics[width=0.48\linewidth]{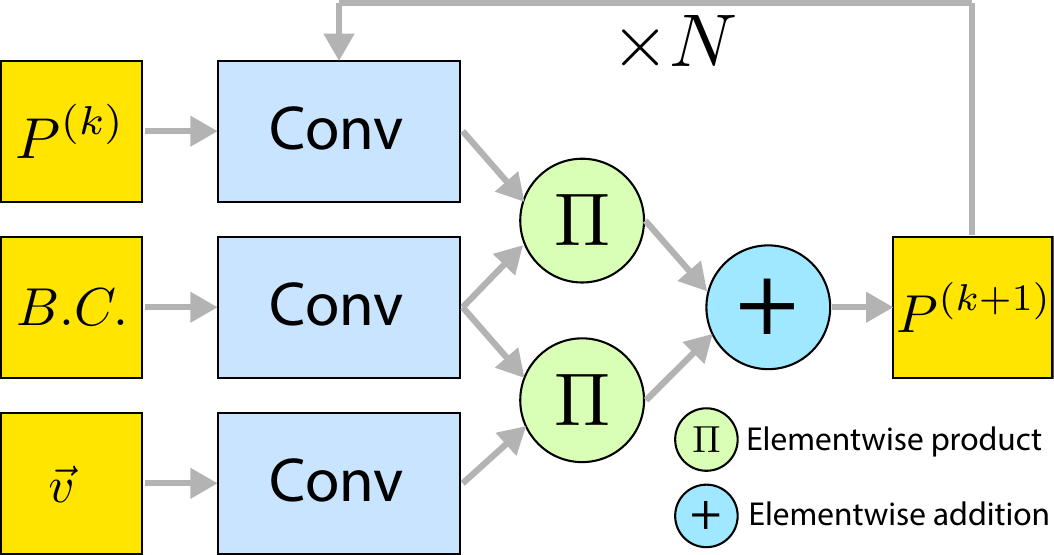}
    \includegraphics[width=0.5\linewidth]{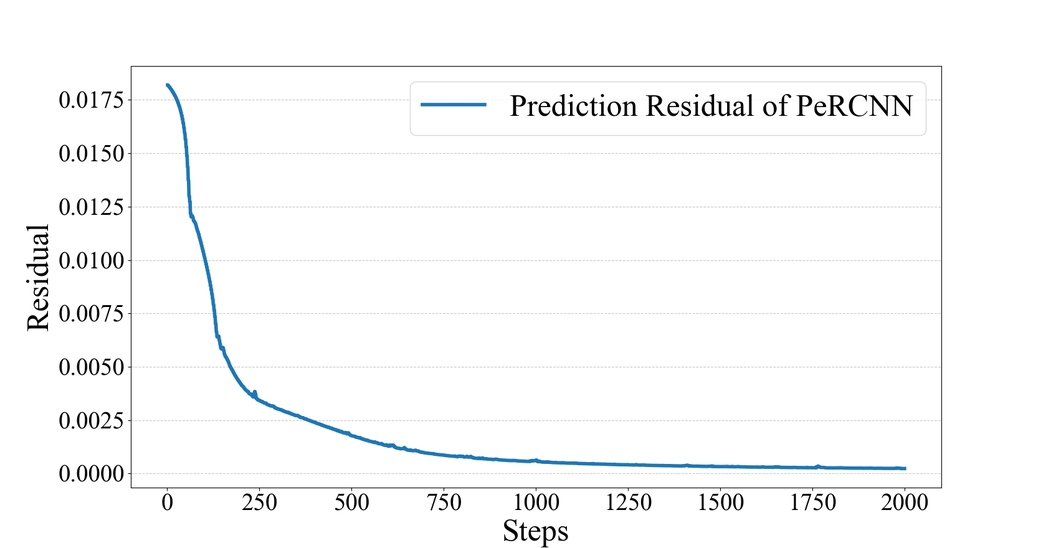}
    \caption{Left: PeRCNN-based optimization network structure for differentiable pressure solving. Right: convergence of the convolutional kernel fitting for the pressure Poisson equation.}
    \label{fig:percnn_str_fit}
\end{figure}

As demonstrated by PDE-Net~\cite{long2018pde}, convolutional kernels with appropriately designed weights can discretize spatial differential operators (gradient, divergence, Laplacian) equivalently to finite-difference stencils. PeRCNN extends this idea with an outer recurrent structure that emulates the iterative process of implicit solvers: at each recurrent layer~$k$, the latent pressure $P^{(k)}$, boundary conditions, and velocity are updated through physics-informed convolutions (Fig.~\ref{fig:percnn_str_fit}, left).

Because the entire forward pass consists of standard differentiable operations (convolutions, activations, element-wise arithmetic), gradients of the pressure solution with respect to input velocity and physical parameters are obtained automatically through backpropagation. This enables end-to-end training without requiring 
analytical gradients of the implicit solver, which sometimes does not align well with forward computation due to numerical approximations.

Prior to deployment, we validate the PeRCNN's numerical accuracy by fitting its convolutional kernel weights and bias to match a second-order Laplacian discretization scheme. Paired data are generated from our reconstructed velocity field, where the pressure before and after solving by a standard implicit solver provide training targets. As shown in Fig.~\ref{fig:percnn_str_fit} (right), the fitted kernel achieves successful convergence. 
Once fitted, the network is reused across all subsequent optimization batches for the same scene without refitting.

\subsubsection{Differentiable Convection}
\begin{figure}[!t]
    \centering
    \includegraphics[width=\linewidth]{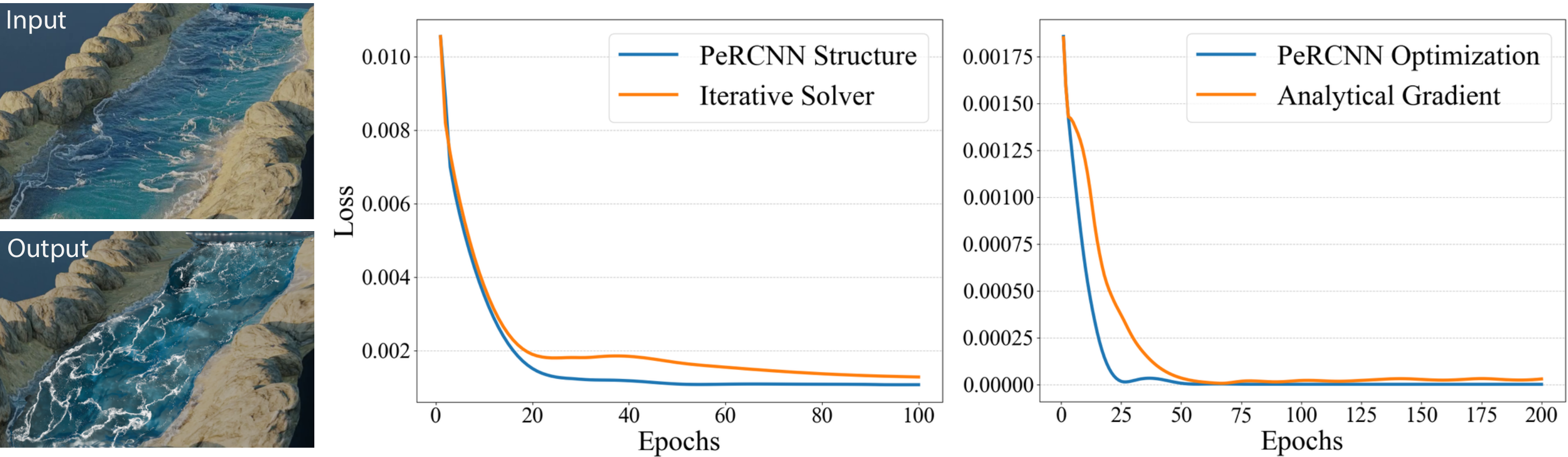}
    \caption{
    We compared the optimization process whose Poisson pressure equation is solved by PeRCNN structure with those solved by (left) Jacobi iteration and (right) analytical gradient. The latter two methods are apt for different batch sizes (analytical gradient degrades for larger batch).
    }
    \label{fig:percnnresult}
\end{figure}

The original APIC method handles convection on particles rather than the grid. To maintain a fully grid-based differentiable evolution, we integrate the semi-Lagrangian back-tracing scheme from stable fluids~\cite{stablefluid99} for unconditional stability. However, back-tracing involves grid indexing and rounding operations that are not smoothly differentiable. We address this by using the discretized gradient of  convection term from the split PDE:
$
    \frac{\partial \mathbf{v}}{\partial t} = - \mathbf{v} \cdot \nabla \mathbf{v}, \quad
    \frac{\partial \mathbf{v}}{\partial v_{i}'} = -\Delta t \, \frac{\partial \mathbf{v}}{\partial x_i},
$
where $\mathbf{v}$ is the post-convection grid velocity and $v_i'$ is the pre-convection velocity in dimension $i$. We employ the upwind scheme for first-order spatial discretization to enhance stability.
Fig.~\ref{fig:percnnresult} compares our PeRCNN-based pressure solver with traditional Jacobi iteration and analytical gradients. The PeRCNN approach achieves comparable accuracy while exhibiting faster convergence during optimization.

\subsubsection{Loss Design and Parameter Optimization}
\label{sec:sim}
With the differentiable grid velocity evolution in place, we optimize the simulation parameters $\Theta = \{\mathbf{v}_{\text{in}}, \tilde{\mathbf{v}}_{\text{in}}, \mathbf{v}_{\text{out}}, \rho, \nu, (b,d)|_{\text{B.C.}},$ $\mathbf{g}, \Delta t\}$. They include bulk inlet and outlet velocities ($\mathbf{v}_{\text{in}}, \mathbf{v}_{\text{out}}$), inlet velocity fluctuations $\tilde{\mathbf{v}}_{\text{in}}$, fluid density $\rho$ and kinematic viscosity $\nu$, boundary bouncing $b$ and damping $d$ coefficients, gravity $\mathbf{g}$, and the simulation time step $\Delta t$. Among them, $\mathbf{v}_{\text{in}}$, $\tilde{\mathbf{v}}_{\text{in}}$, $\mathbf{v}_{\text{out}}$,  and $(b,d)|_{\text{B.C.}}$ are field properties defined on the detected inlet/outlet grids or terrain grids, as in Fig.~\ref{fig: grid_bc}. Since the geometry is generated rather than metrically reconstructed, $\Theta$ holds effective simulation parameters of the reconstructed scene rather than uniquely identifiable constants of the real fluid.

The optimization is under the $L_2$ loss of velocity magnitude difference and extra directional alignment, applied over a grid mask distinguishing between surface and occlusion, kinetic and stationary, occupied and empty cells:
\begin{equation}
\begin{aligned}
    \min_\Theta ~\text{gridMask} \circ \bigg( &\alpha \Big(1-\frac{\mathbf{v}_{\text{sim}}}{|\mathbf{v}_{\text{sim}}|}\cdot\frac{\mathbf{v}_{\text{ref}}}{|\mathbf{v}_{\text{ref}}|}\Big)+ \beta \,L_2(\mathbf{v}_{\text{sim}},\mathbf{v}_{\text{ref}})\bigg),\\
    \mathbf{v}_{\text{sim}}^{i+n} = \text{DiffSim}^n &\big(\mathbf{v}^{i}_{\text{ref}};\;\Theta\big) 
\end{aligned}
\label{eq::optLoss}
\end{equation}
where $\mathbf{v}_{\text{ref}}$ denotes the reference velocity field from the previous stage. Each training sample is within the $N$-frame window (Sec.~\ref{sec::unionn}) up to 25 continuous simulation steps, with approximately 80 recurrent layers of PeRCNN per step.
We set $\alpha=0.67$ and $\beta=0.33$. Physical and image terms are not weighed against each other here, since the video enters through $\mathbf{v}_{\text{ref}}$ and gridMask, both already shaped by the constraints of Sec.~\ref{sec:surface-motion} and Sec.~\ref{sec:volumetric}.

To mitigate gradient vanishing in this deep graph, we adopt several strategies: 
(1)~reasonable initialization (proper scale of in/out velocity, non-zero gravity, CFL-satisfying time step); 
(2)~upwind scheme to compute gradient of the convection term; 
(3)~dynamically adjusted window size $N$ (longer for steady-state fluids, shorter for rapid flows); 
and (4)~Parameters are re-parameterized via domain-specific activations (exponential for positive scales and sigmoid for bounded values) to enforce physical constraints and ensure stable gradient flow across different magnitudes.

\subsubsection{Forward Simulation and Gaussian Coupling}
With the optimized parameters $\Theta^*$, we re-simulate the fluid using the full APIC method, coupling Lagrangian particle advection with Eulerian grid dynamics.
The simulation properties and Gaussian properties are in two-way coupling: Gaussian points provide the initial positions and velocities for simulation particles, while the computed affine matrix $\mathbf{C}$ records local deformation and updates the Gaussian covariance:
\begin{equation}
\begin{aligned}
    \mathbf{F}_p^{n+1} &= (\mathbf{I}+\Delta t \,\mathbf{C}_p^{n+1})\mathbf{F}_p^n, \\
     \mathbf{A}_p^n &= \mathbf{F}_p^n \mathbf{A}_p^0 \mathbf{F}_p^\mathsf{T}.
\end{aligned}
\end{equation}
The affine matrix $\mathbf{C}$ approximates the local velocity gradient. The deformation gradient $\mathbf{F}$ is updated at each time step, deforming the Gaussians as reflected in the evolution of their covariance $\mathbf{A}$.

\begin{figure}[!t]
    \centering
    \includegraphics[width=\linewidth]{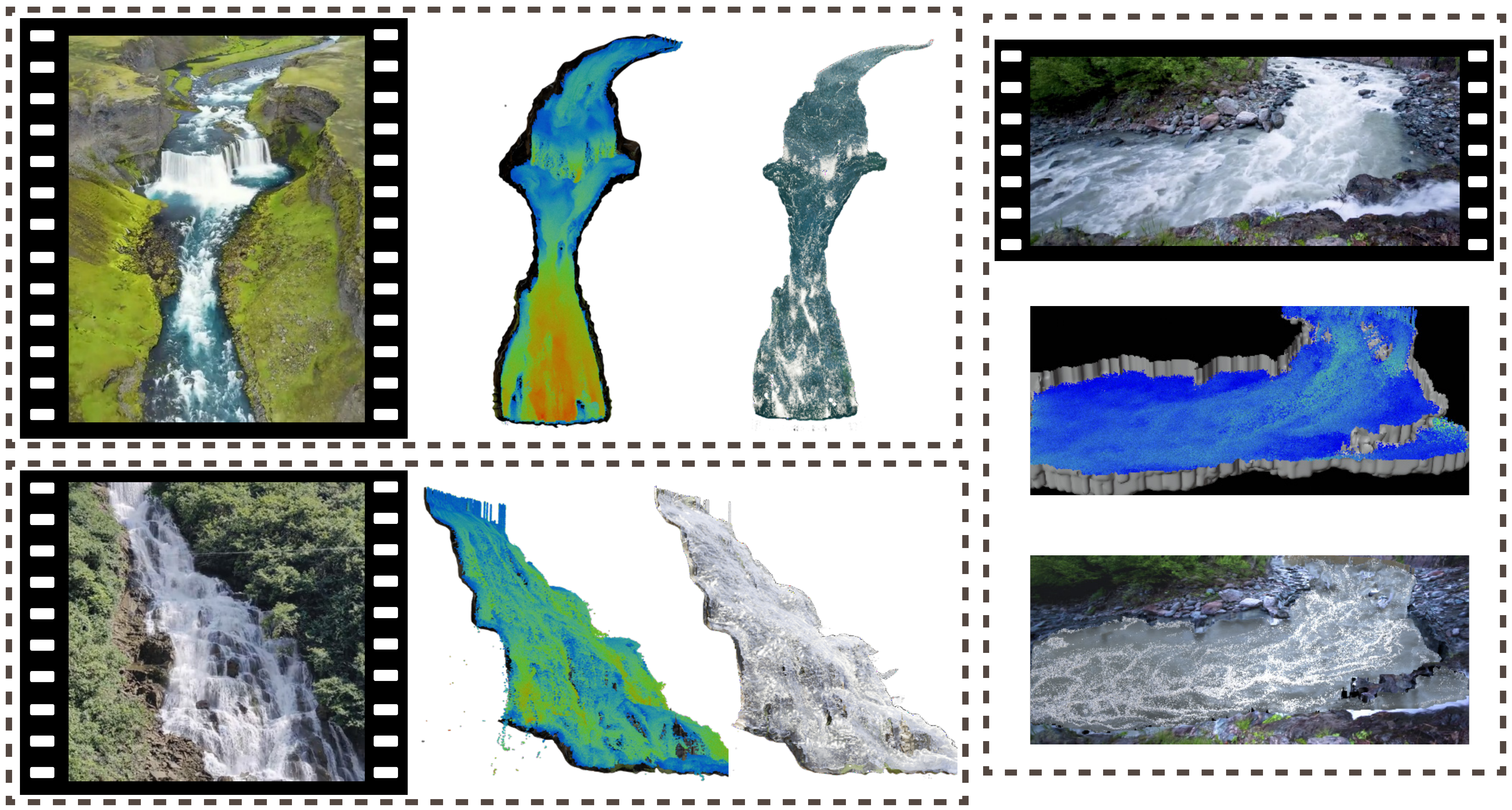}
    \caption{In each scene, the middle figure shows the re-simulation of our final output geometry and physics based on the input video on the left/top, which is then rendered in the figure on the right/bottom.
    }
    \label{fig: raw}
\end{figure}

An example of the optimized effect is shown in Fig.~\ref{fig: raw}.

\subsection{Result Rendering}
The output dynamic 3D Gaussian assets can be rendered in two complementary modes, reflecting the viscous/inviscid distinction discussed in the introduction.

\textbf{Direct splatting} retains the originally generated spherical harmonic features and renders via standard 3DGS rasterization. It suits \textit{viscous fluids} (e.g., honey, lava) whose persistent per-Gaussian appearance lets the optimized appearance evolve coherently with the motion, and fluids with uniform texture benefit likewise. It cannot, however, interact with environmental lighting, limiting its use in composed digital scenes.

\textbf{Relighting-based rendering} suits \textit{inviscid fluids} (e.g., flowing rivers), where the spherical harmonic features of newly injected inlet particles are unknown and unoptimized interior particles become exposed, making initial-frame appearance unreliable.
\textit{Splatting-based relighting} adapts Gaussian relighting frameworks~\cite{jiang2024gaussianshader} with transparent light absorption~\cite{feng2025splashing}, reaching real-time speeds on modern GPUs. It is, however, incompatible with mesh-based rendering pipelines, and its lighting interaction is limited to surface reflection with a model fitted during training, so it cannot respond to arbitrary environmental lighting at runtime.
\textit{Mesh-based rendering} offers more flexible and advanced illumination by reconstructing meshes from the Gaussian particles. We then apply standard illumination pipelines compatible with the target scene. The post-generation optimization (Sec.~\ref{sec: postopt}) is particularly beneficial here, as it produces more densified, isotropic, and uniform Gaussians that yield better alignment between dynamics and appearance.
For base color in both relighting methods, we homogenize the zero-order spherical harmonic coefficients, providing temporally coherent surface appearance even with inflow particles.

\begin{figure}[!t]
  \centering
  \includegraphics[width=\linewidth]{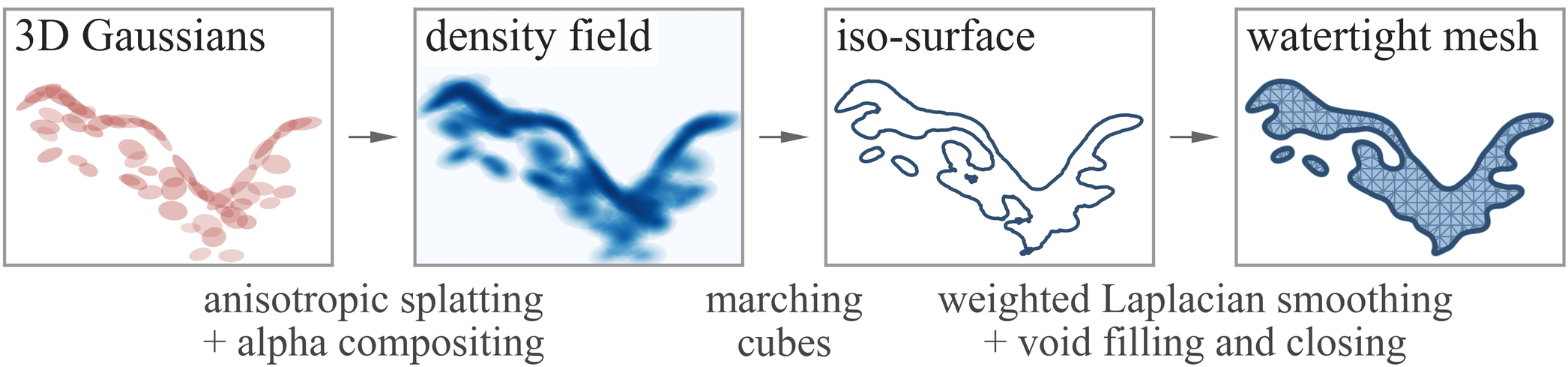}
  \caption{%
    From 3D Gaussians to a watertight mesh, shown in cross-section.
    Alpha-compositing the kernels gives a density field continuous across
    inter-primitive gaps, so the extracted level set is already closed.%
  }
  \label{fig:gs2mesh}
\end{figure}

Figure~\ref{fig:gs2mesh} shows the steps. We splat each Gaussian with its own
anisotropic kernel, truncated at $2\sigma$, and alpha-composite them into a
density field $1-\prod_p\big(1-\sigma_p e^{-m_p^2/2}\big)$, with $m_p$ the
Mahalanobis distance. The kernel support is wider than a voxel, so the field is
continuous across inter-primitive gaps and its level set is closed by
construction. Marching cubes, sliver removal and a weighted Laplacian pass then
follow~\cite{loschner2023weighted}. Its weights spare the sparse regions, so
droplets keep their volume. We then clear the remaining spurious topology on the occupancy
rather than the mesh, filling interior cavities, closing the cracks and discarding
components below a size threshold.

\begin{figure}[t]
    \centering
    \includegraphics[width=0.9\linewidth]{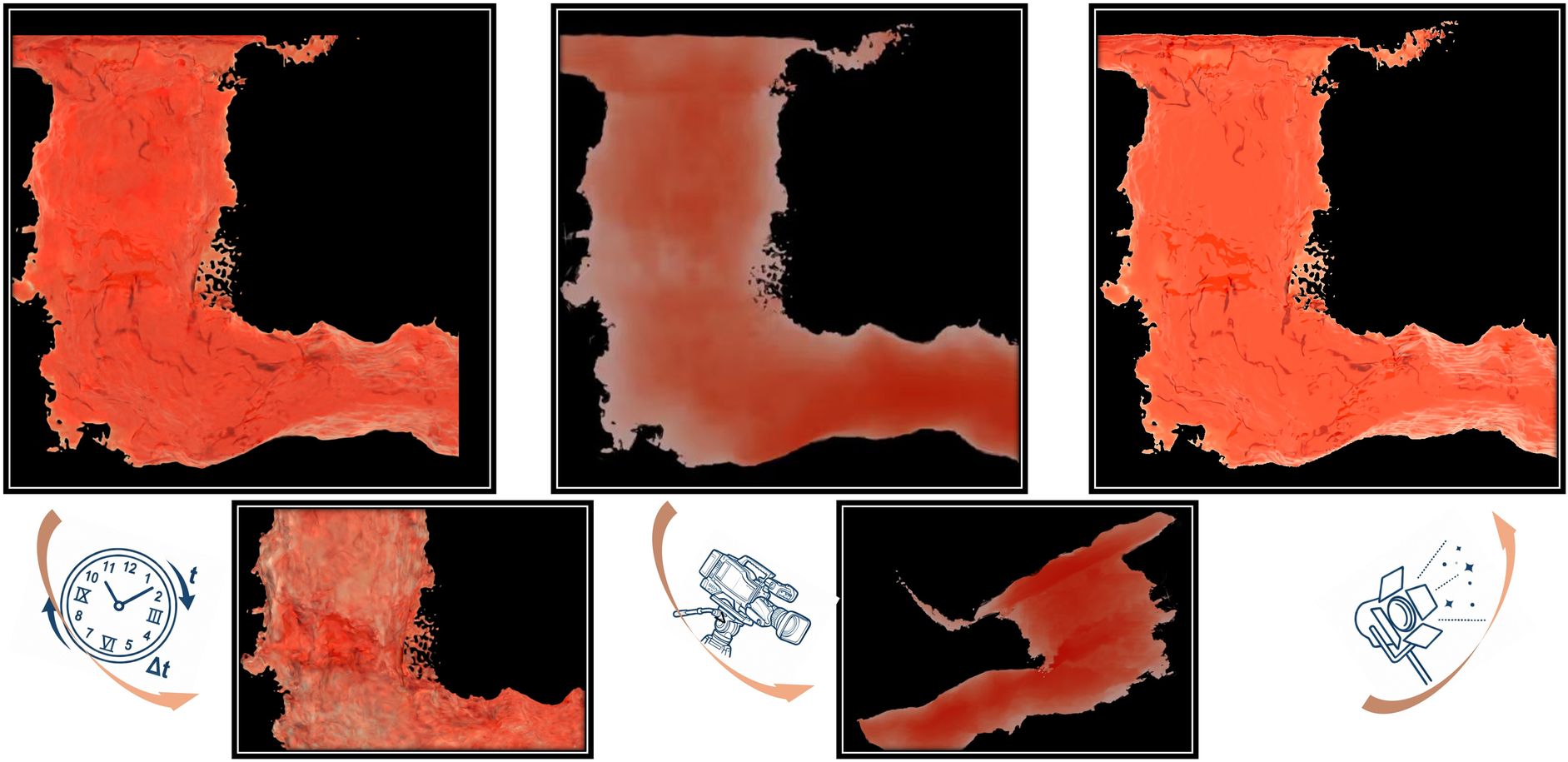}
    \caption{Comparison of rendering options. Left: direct splatting preserves per-Gaussian appearance but degrades as interior particles are exposed over time. Middle: splatting with volumetric light absorption and homogenized color achieves temporal and multi-view consistency, as the absorption conveys depth perception under varying camera angles. Right: surface re-meshing enables specular light reflection.}
    \label{fig:rendop}
\end{figure}

Figure~\ref{fig:rendop} illustrates these rendering options. Single-view optimization fits the 3DGS features to the current frame quite accurately, but as interior particles become exposed during simulation, the appearance grows chaotic. Using a modified Gaussian kernel with volumetric light absorption and homogenized fluid color achieves both temporal and multi-view consistency, though without specular light reflection. Re-meshing the surface with smoothing enables the fluid to respond to environmental lighting, producing more realistic illumination effects.

\subsection{Editability and Interaction}
Because our output contains optimized simulation parameters rather than fixed per-frame deformations, the resulting asset is inherently editable and interactive.
These are intuitive quantities that artists can modify without specialized simulation expertise, and any modification is propagated through forward re-simulation, ensuring physically consistent dynamics.
Compared to canonical and displacement field fitting, which possess high degrees of freedom and are thus difficult to interpret or modify in a controlled manner, our compact parameterization provides semantically meaningful control over fluid behavior.

Users can alter the fluid material (density/viscosity), add interacting solids, extend the simulation duration, or change viewing directions, all automatically achieved by simulation with updated parameters or scene geometry.
Moreover, the output also includes the underlying terrain geometry, providing enriched context for integration into existing 3D scenes and post-processing applications such as terrain editing.

\section{Results}

Our framework is implemented mainly in \texttt{Python} on the workstation equipped with a 24-core \texttt{Intel Xeon Platinum 8255C} CPU and an \texttt{NVIDIA Tesla V100} GPU, or \texttt{NVIDIA RTX 5090}.
For 3DGS generation (Sec.~\ref{sec:geometry}), we port the open-source TriplaneGaussian~\cite{zou2024triplane} implementation (used by default) and retain intermediate outputs such as estimated camera parameters and segmentation parameters for subsequent processes. The appearance optimization builds upon the original 3DGS codebase~\cite{kerbl3Dgaussians}.
Surface motion extraction and volumetric velocity construction (Sec.~\ref{sec:surface-motion},~\ref{sec:volumetric}) are implemented with OpenCV and OpenGL utilities, where parallelizable operations are accelerated via \texttt{Taichi}~\cite{hu2019taichi}.
The differentiable APIC simulation and parameter optimization (Sec.~\ref{sec:optimization}) is also GPU-parallelized with \texttt{Taichi}. The simulated 3D Gaussians are compatible with standard 3DGS visualizers. Splatting relighting is realized through modified CUDA kernels, and mesh-based rendering leverages the surface reconstruction of \texttt{splashsurf}~\cite{loschner2023weighted}.

\begin{figure}[!t]
    \centering
    \includegraphics[width=\linewidth]{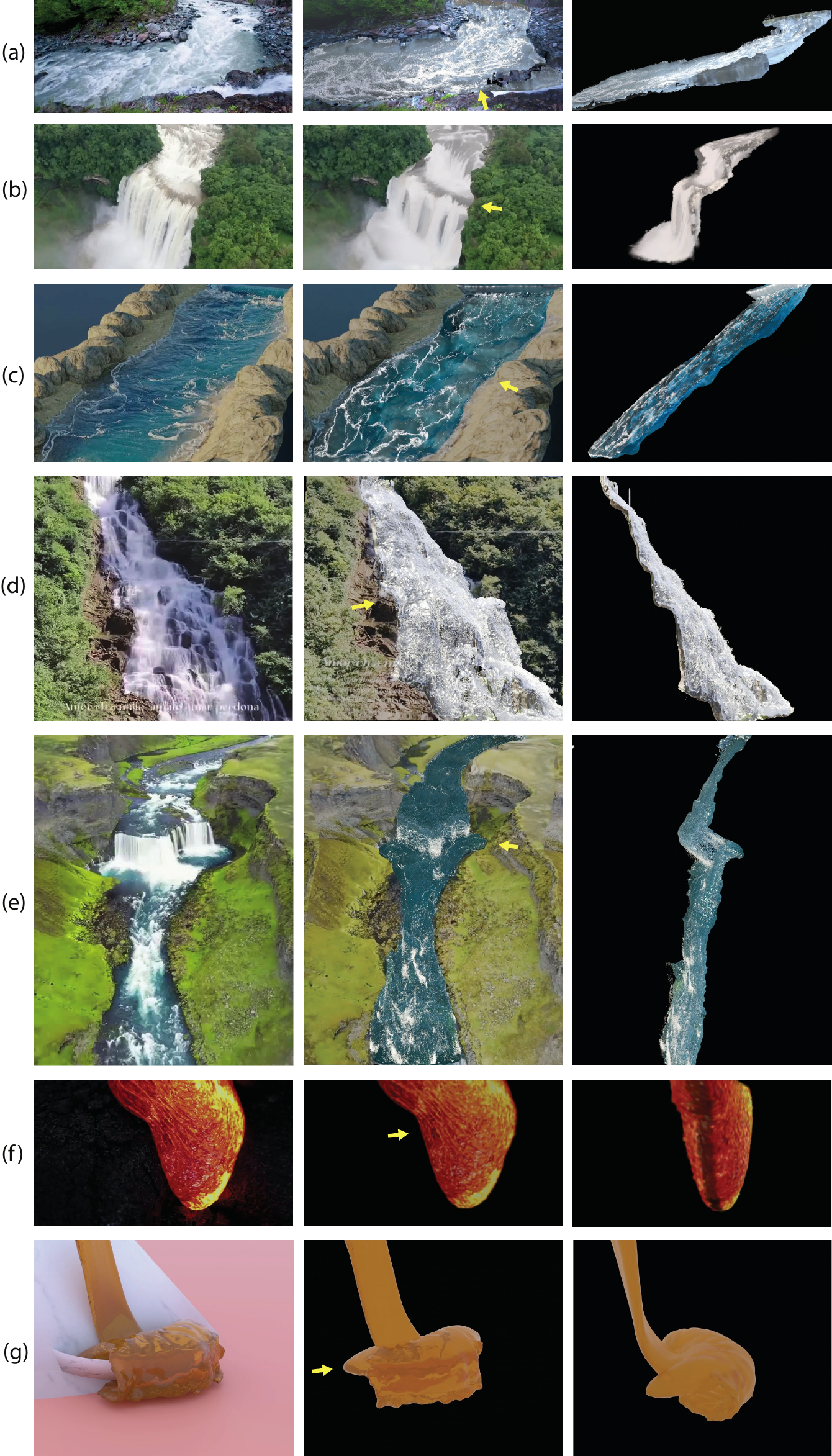}
    \caption{Examples of our output. Left column: frames from the input videos. Middle column: frames from the rendered output. Right column: side views of the output fluid, captured from camera angles indicated by the yellow arrows in the middle images.
    The last two rows correspond to viscous fluids and they, together with (b),  are directly rendered by splatting.}
    \label{fig:all}
\end{figure}

\subsection{Parameter Optimization and Velocity Evaluation}
\begin{figure}[!t]
    \centering
    \includegraphics[width=\linewidth]{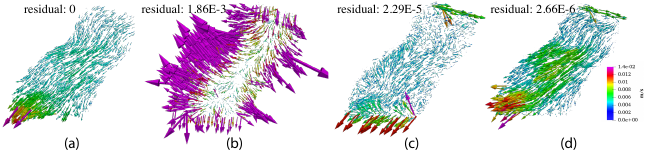}
    \caption{
    Volumetric velocity fields: (a) reference field extracted from the video, (b) simulation with initialized parameters before optimization, (c) optimized with $L_2$ loss only, and (d) optimized with full loss terms. Arrow color and size indicate velocity magnitude. The residual above each panel reports the prediction loss against (a).
    }
    \label{fig:3dvcomp}
\end{figure}

Figure~\ref{fig:3dvcomp} compares four volumetric velocity fields $\textbf{V}_{vol}$ for the scene in Fig.~\ref{fig:all}~(c): a reference field extracted from the input video (Sec.~\ref{sec:surface-motion},~\ref{sec:volumetric}), a forward simulation with properly initialized physical parameters, and simulations with optimized parameters using either partial or full loss terms as defined in Eq.~\ref{eq::optLoss}.
This comparison serves as an ablation study verifying both the necessity and efficacy of the parameter optimization. Without it, even using the same reconstructed geometry, the simulation produces physically implausible over-flowing and over-accelerated patterns (b). 
Optimizing with the $L_2$ loss alone (c) reduces the residual but tends to suppress velocities toward a near-steady state, since low acceleration trivially reduces the loss
to an adequate level (recover main stream flows) but loses finer details.
Adding directional alignment (d) corrects this: even when individual velocity components are close, the local flow direction can differ noticeably, and the directional term penalizes such misalignment, producing a field both numerically and visually consistent with the reference.
Meanwhile, the optimized simulation can also correct optic flow deficiency present in the video-extracted reference field, since the simulation enforces physical laws.

Table~\ref{table:1} compares scale of representative parameters and the simulation outputs before and after the optimization.
Starting from initializations at plausible scales, the optimizer consistently converges to stable values.
This convergence is expected, 
as these values represent the optimum for the given reconstructed terrain and the estimated velocity field.
In contrast, simulations with unoptimized parameters overflow and diverge even on the same geometry (Fig.~\ref{fig:3dvcomp}~(b)), and small discrepancies accumulate over successive steps even when individual parameters start at comparable scales. Sec.~\ref{sec::opt_necessity} examines this in detail.
\begin{table}[!t]
\caption{Representative simulation parameters and outputs before and after optimization. Geometry scale follows Fig.~\ref{fig:3dvcomp}.}
\vspace{-10pt}
\label{table:1}
\centering
$$
\setlength{\arraycolsep}{2.5pt}
\begin{array}{ccccccc}
\noalign{\hrule height 1.5pt}
    \multicolumn{3}{c}{\textbf{Physical Simulations}} & & \multicolumn{3}{c}{\textbf{Non-uniform Parameters}} \\
    \cline{1-3} \cline{5-7}
    \vec{v}_{3D} \scriptstyle[m/s] &  \times 10^{-1} &  \times 10^{-3} & & \vec{v}_{in} \scriptstyle[m/s] & \times 10^{-2} & \times 10^{-3} \\
    p\scriptstyle[Pa] &  \times 10^{1} &  \times 10^{-3} & & \vec{v}_{out}\scriptstyle[m/s] & \times 10^{-2} & \times 10^{-3} \\
    \cline{1-3}
    \multicolumn{3}{c}{\textbf{Uniform Parameters}}  & & bouncing \scriptstyle[] & \times 10^{-1} & \times 10^{-1}\\
    \cline{1-3}
    \vec{g}\scriptstyle[m/s^2] & \times 10^{-2} & \times 10^{-3} &  & & \text{before opt.} & \text{after opt.} \\
    \cdashline{5-7}
    dt\scriptstyle[s] & \times 10^{-2} & \times 10^{-1} & &
    \multicolumn{3}{c}{\text{geometry scale : }  ~10^{-2} ~\scriptstyle m} \\
    \noalign{\hrule height 1.5pt}
\end{array}
$$
\end{table}

\begin{figure}[!t]
    \centering
    \includegraphics[width=\linewidth]{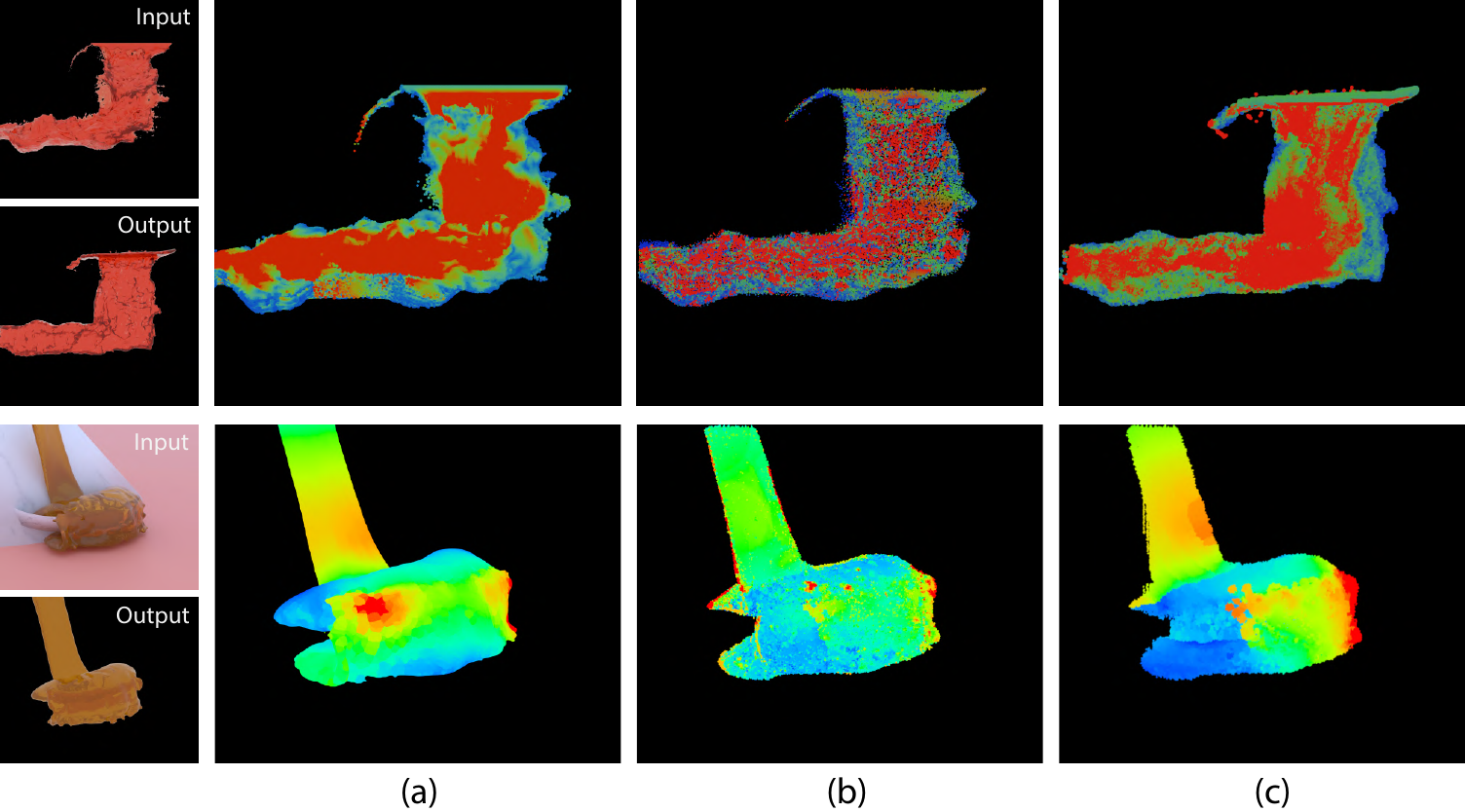}
    \caption{Planar velocity comparison on synthetic scenes: red river (top) and honey (bottom); (a)~Ground-truth planar velocity , (b)~estimated velocity from surface motion extraction, and (c)~velocity field after full parameter optimization. Color encodes the  magnitude.}
    \label{fig:vcomp}
\end{figure}

To quantitatively assess velocity accuracy, we made two synthetic test videos from known fluid simulations (a river and a honey pour) to compare the recovered velocity fields against ground truth. Because our pipeline operates on inter-frame screen-space displacements (discretized by frame rate and scaled to screen dimensions), both the ground-truth and recovered 3D velocities are projected into 2D NDC space for a fair comparison.
Figure~\ref{fig:vcomp} presents the results. The velocity estimated from surface motion extraction and volumetric construction (column~b) captures mesoscale dynamics but retains noise on fluid surfaces. After parameter optimization (column~c), the simulation produces smoother and more physically coherent fields, recovering details such as boundary-induced deceleration in the river flow and strain development in the honey's inflow region.

\subsection{Stage-wise Validation of Velocity Field Extraction}
\label{sec::stagewise}

\begin{figure}[!t]
  \vspace{-10pt}
  \centering
  \includegraphics[width=0.9\linewidth]{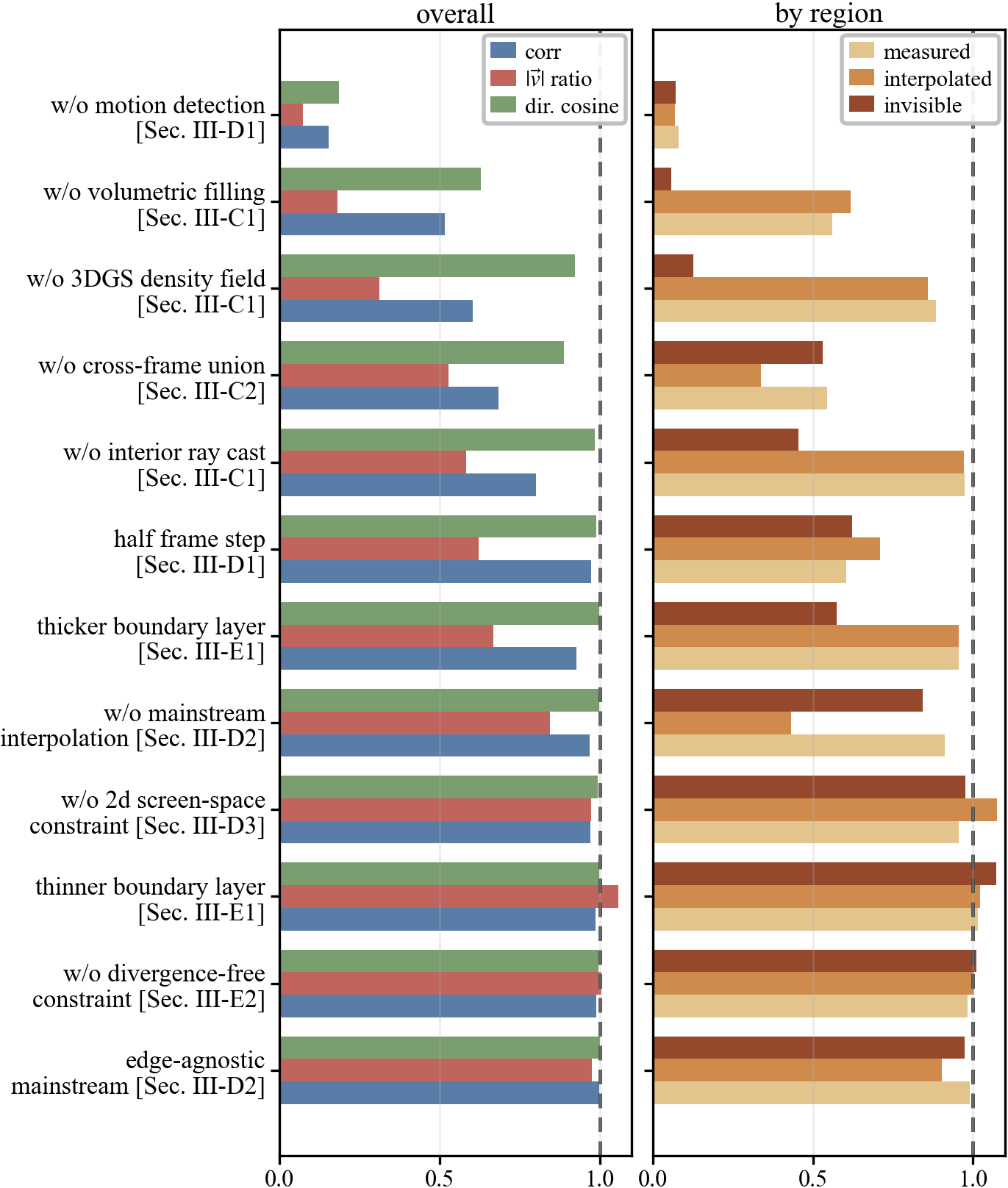}
  \vspace{-10pt}
  \caption{%
    Stage-wise ablation, as ratios to the full pipeline (dashed line). Left:
    correlation, mean speed and median direction cosine over all fluid cells.
    Right: the speed ratio by how each cell obtained its velocity.%
  }
  \label{fig:stageablation}
\end{figure}

Figure~\ref{fig:stageablation} ablates the velocity extraction chain one stage at
a time on the scene of Fig.~\ref{fig:all}~(c), comparing each volumetric field
against the complete pipeline over the same fluid cells. Correlation, mean speed and median direction cosine
isolate errors in structure, magnitude and direction.
The extraction chain is not uniformly sensitive. Removing screen-space motion
detection leaves essentially nothing ($\text{corr}\!=\!0.15$), and the geometry stages
each reduce correlation by $0.3$ to $0.5$, while the refinements applied to an
already-populated field by less. The right panel splits the speed ratio by how
each cell
obtained its velocity. Most ablations degrade their own stage's region while
sparing the others, and withholding volumetric filling leaves the observed
region at $0.56$ but the unseen interior at $0.06$. Since $77\%$ of the fluid is
never observed, these stages make the output volumetric rather than a surface
field.

\begin{figure*}[!t]
  \centering
  \includegraphics[width=0.9\textwidth]{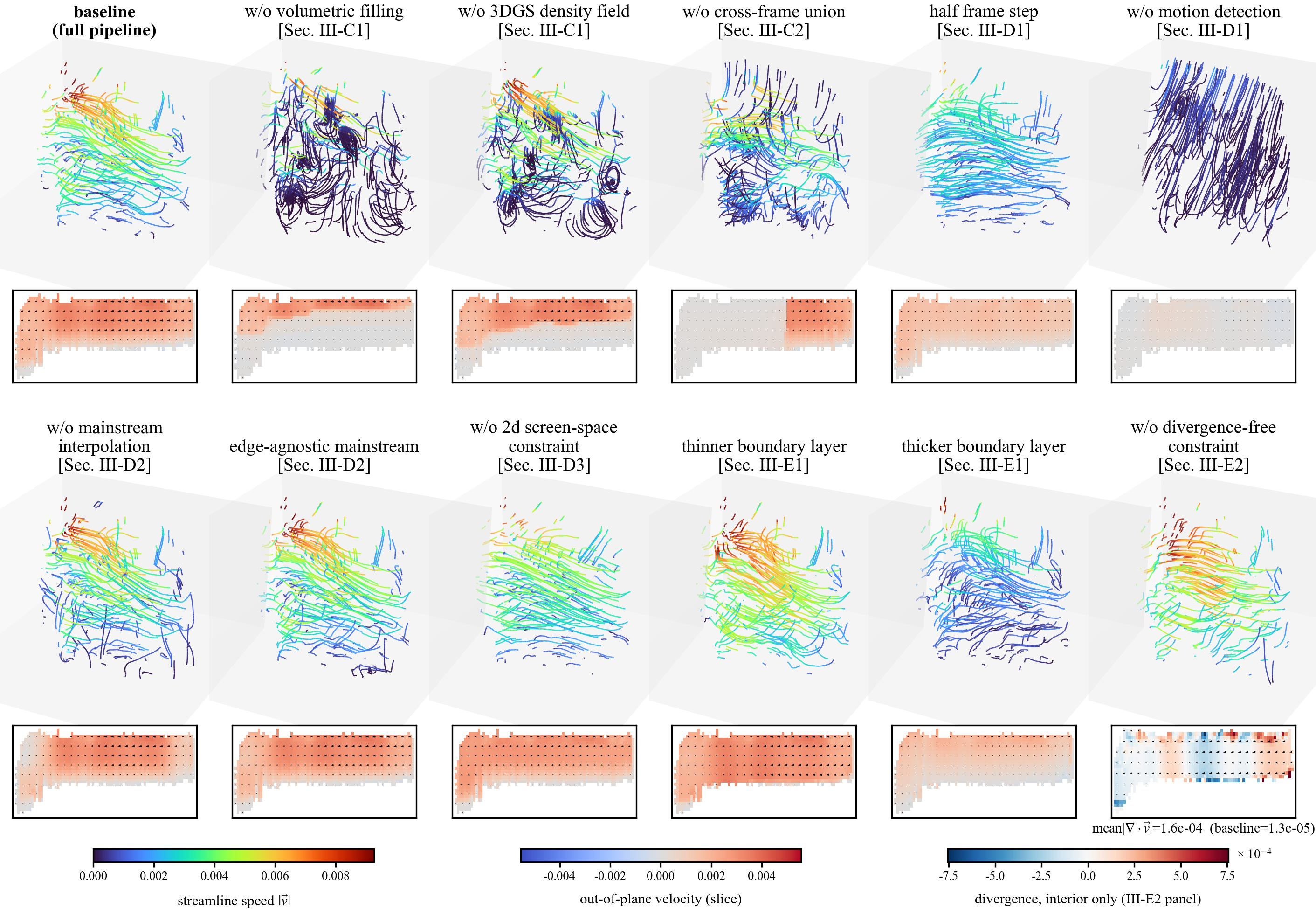}
  \caption{%
    Stage ablations in 3D. Streamlines share seed points and color scale; below
    each, a mid-depth slice with in-plane velocity as arrows and out-of-plane as
    color. The last panel shows divergence instead.%
  }
  \label{fig:stage3d}
\end{figure*}

\begin{figure}[!t]
  \centering
  \includegraphics[width=0.83\linewidth]{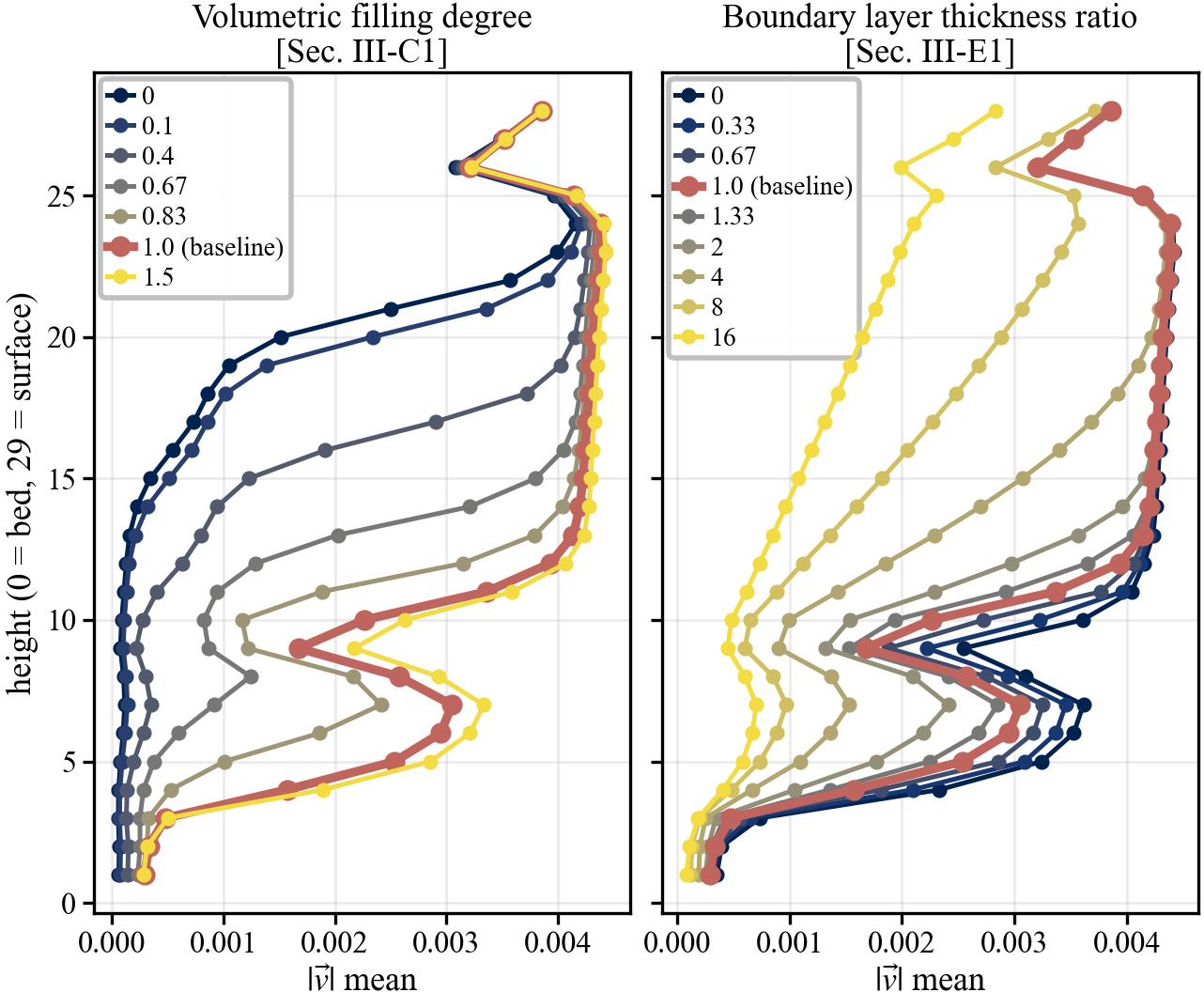}
  \caption{%
    Mean speed against height for the filling degree (left) and boundary-layer
    thickness (right), as ratios to their defaults (red).%
  }
  \label{fig:stagedepth}
\end{figure}

Figure~\ref{fig:stage3d} renders the same ablations as streamlines from
identical seed points. Fields that lose their interior produce short, broken
streamlines below. Fig.~\ref{fig:stagedepth} resolves two stages by height. The
filling degree traces a saturation curve towards our default, while the boundary
layer damps the velocity as it thickens, its profiles staying grouped around the
default $\delta$ of Eq.~\ref{eq::wallFn} at middle and upper depths. 
Both responses saturate near the values our method adopts, 
which supports the necessity of these stages and the defaults chosen for them.

\begin{figure*}[!t]
  \centering
  \includegraphics[width=0.9\textwidth]{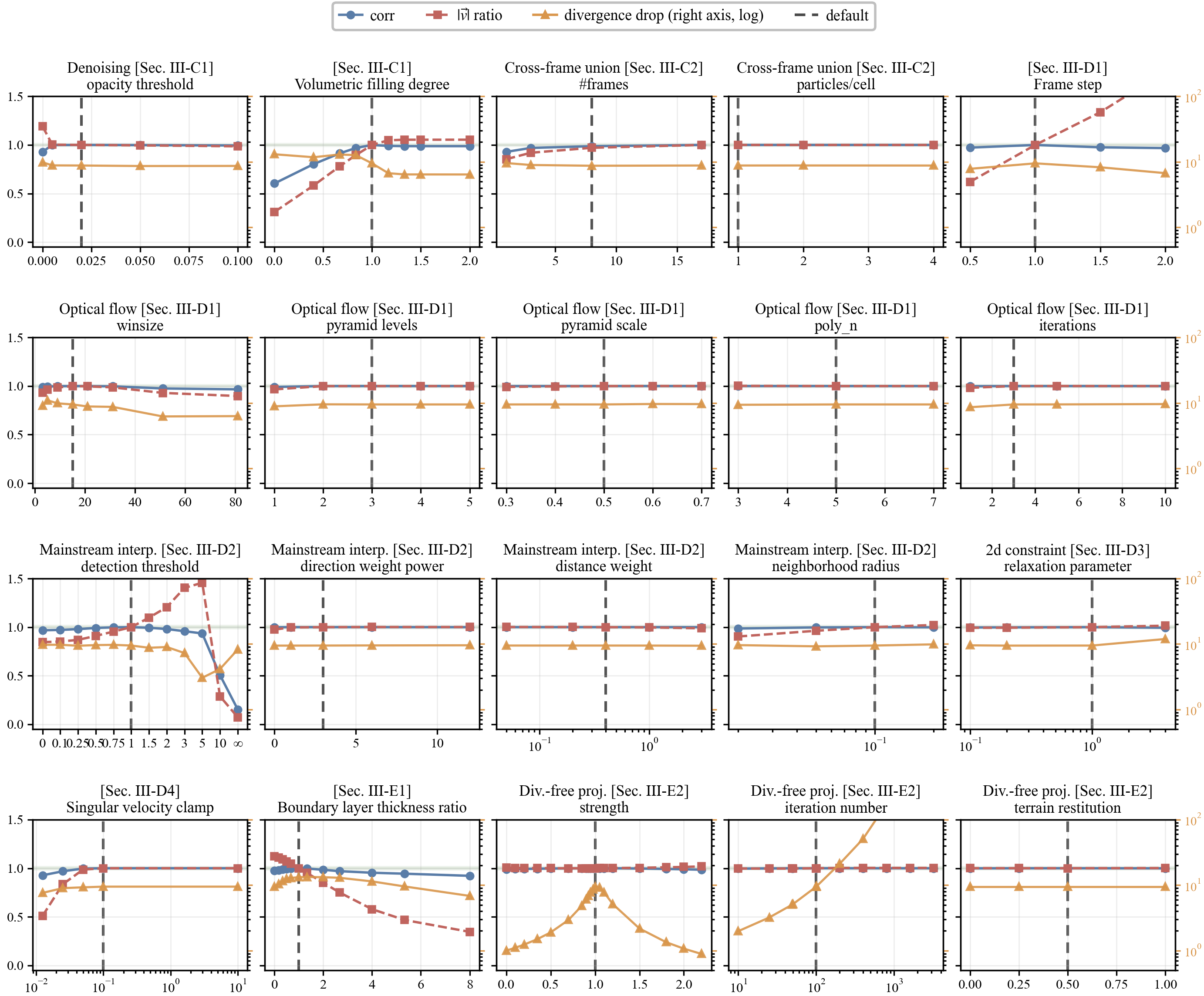}
  \caption{%
    Hyperparameter sensitivity. Each panel sweeps one hyperparameter about its
    default (dashed line): correlation and mean speed against the full pipeline
    (left axis), divergence reduction from the projection (right axis, log). The
    denoising, particles-per-cell and frame-count families use a regenerated
    point cloud as reference.%
  }
  \label{fig:stagesensitivity}
\end{figure*}

Sweeping $20$ hyperparameters (Fig.~\ref{fig:stagesensitivity}), the optical-flow
solver settings are essentially flat, while the quantities our pipeline
introduces each show a plateau rather than a sharp optimum, 
which makes the defaults transferable across scenes without per-scene tuning.
The dependencies
that do appear are expected. Speed grows with the frame step, divergence keeps
falling with projection iterations beyond what is required, and the correction
strength is most effective at the value prescribed by Eq.~\ref{eq::divProj}.
The supplementary material lists main hyperparameters, which can be used across different scenes.

\subsection{Necessity and Robustness of Parameter Optimization}
\label{sec::opt_necessity}

The volumetric field of Sec.~\ref{sec:volumetric} supplies an initialization,
not a parameter set. We examine how the optimization depends on its starting
point, what it recovers, and what each group contributes.

\subsubsection{Robustness to Initialization}
\label{sec::init_robust}

\begin{figure}[!t]
  \centering
  \includegraphics[width=0.9\linewidth]{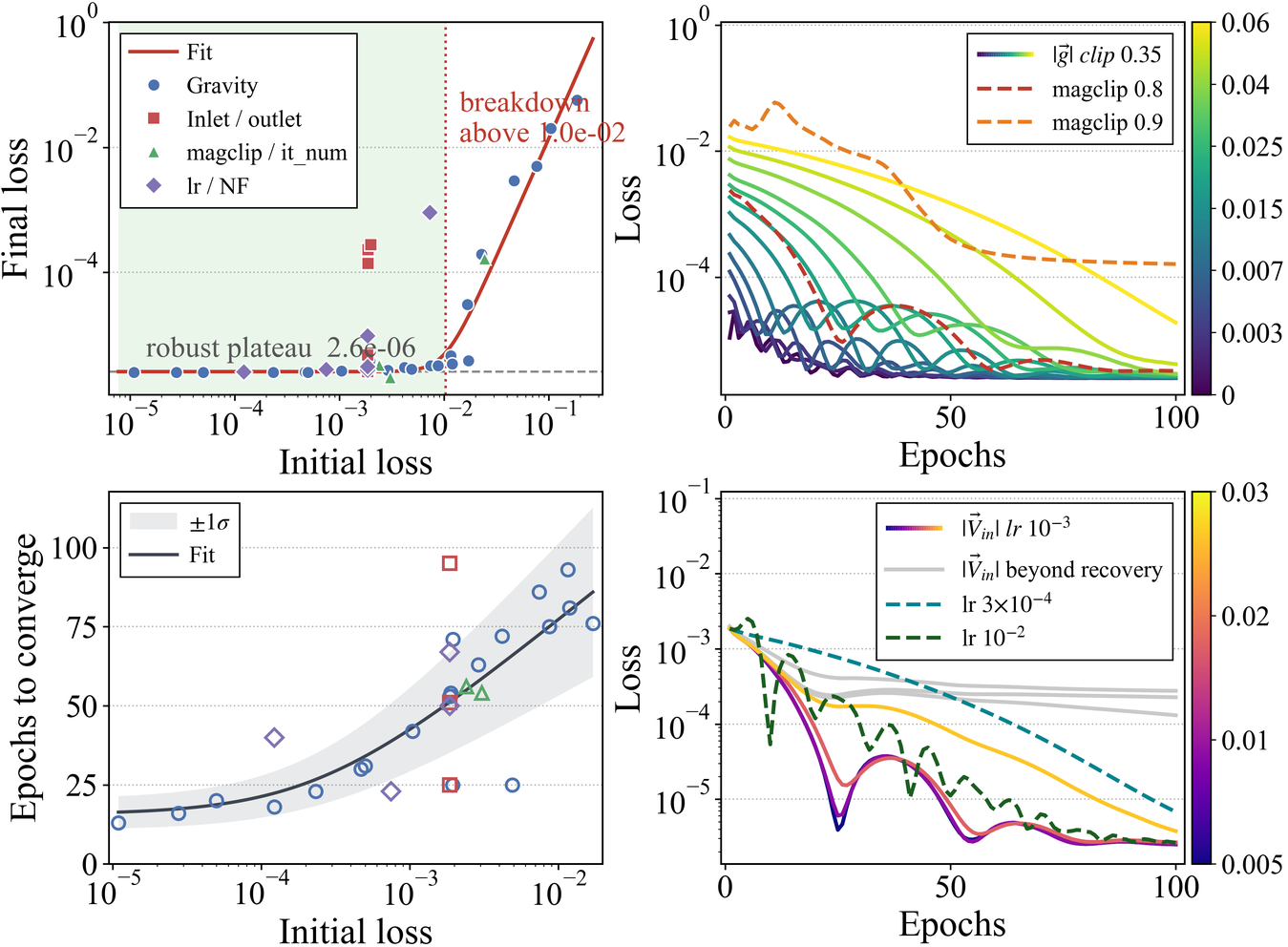}
  \caption{
    Robustness to initialization. Left: final loss and epochs to convergence
    ($\text{loss}<5\times10^{-6}$) against initial loss. Right: loss
    trajectories. Loss is the objective of Eq.~\ref{eq::optLoss}.%
  }
  \label{fig:optrobust}
  \vspace{-10pt}
\end{figure}

Figure~\ref{fig:optrobust} sweeps the initial value of each parameter group
over three orders of magnitude. Nearly every run reaches the same final loss, so
the starting point decides how long the optimization takes rather than what it
converges to. Initial loss is not comparable across groups, since each perturbs the field
differently, so the trend against it is mainly a tendency.
What does break the optimization is a physically implausible start. The grey
trajectories point the inlet out of the scene and never recover, although their
initial loss looks ordinary because the inlet covers few cells. Tabulated values
help exactly here, placing the optimization in the right basin so it proceeds
without further input. Learning rate and clipping change the path but
not the converged solution, except for extreme numerical settings.

\subsubsection{Optimization Trajectory across Scenes}
\label{sec::opt_evolution}

\begin{figure}[!t]
  \centering
  \includegraphics[width=\linewidth]{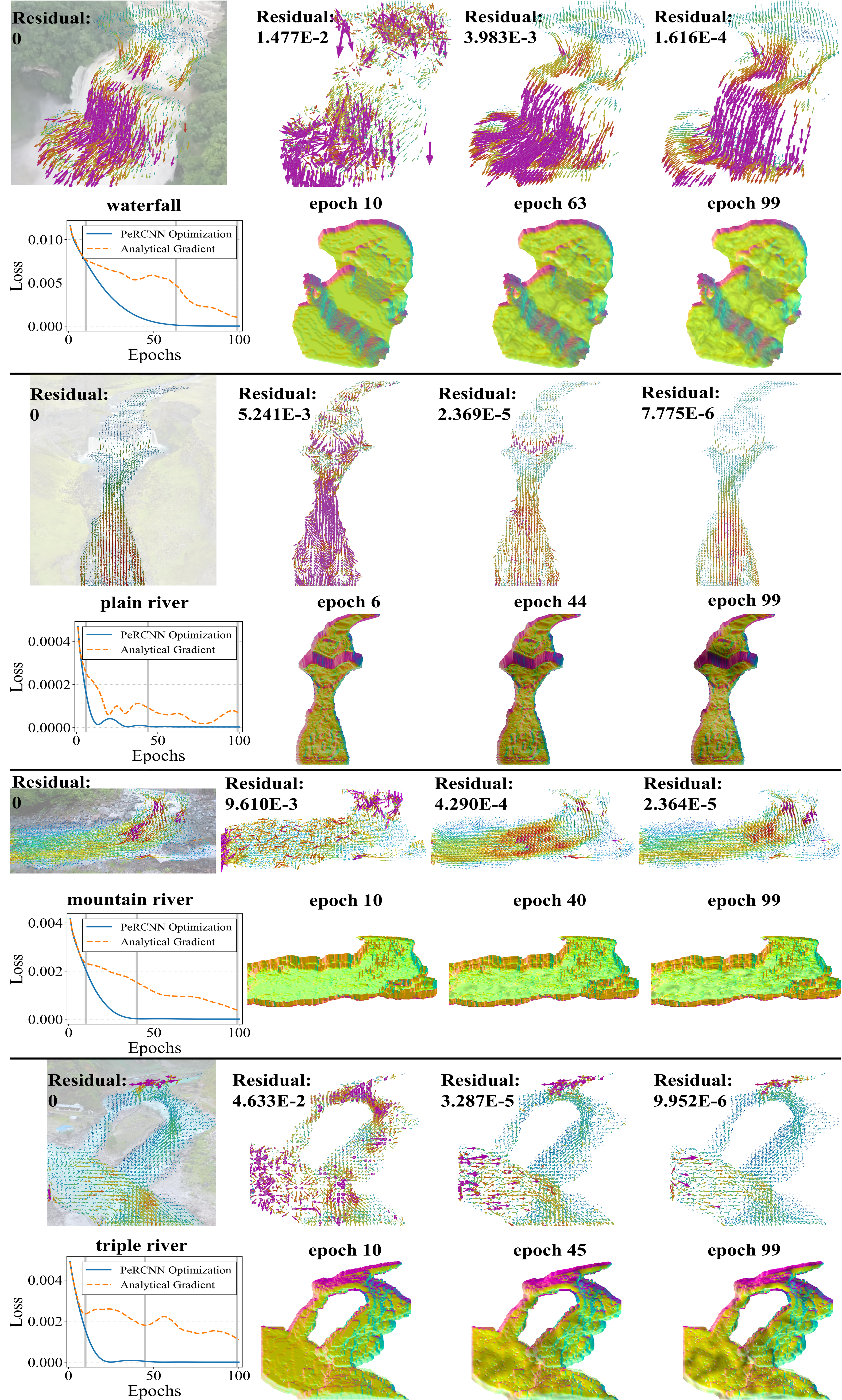}
  \caption{%
    Evolution of the optimized parameters, one block per scene. Left: reference
    field and loss trajectory. Right: simulated velocity (top) and
    $b|_{\text{B.C.}}$ scaled by the terrain normal (bottom), its components
    in camera coordinates as RGB. \emph{Residual} is the measured by mean squared error.%
  }
  \label{fig:optevolution}
\end{figure}

Figure~\ref{fig:optevolution} follows the optimization on four captured scenes.
The terrain normal comes from the generative 3DGS, then the response
coefficients are refined, darkening where the bouncing coefficient decreases. Setting these by hand is not
practical, because $(b,d)|_{\text{B.C.}}$ is a field defined per terrain cell
rather than a single number, as is
$\tilde{\mathbf{v}}_{\text{in}}$. The geometry is also a generation
under perspective transform rather than a metric reconstruction, so values
correct for real water need not reproduce the observed motion. Tabulated constants are
thus a good starting point, but not a substitute for the optimization.
Analytical gradients are cheaper per epoch but reach a higher loss at equal
epochs and are unsteady on some scenes. We use the PeRCNN by default.

\subsubsection{Contribution of Parameter Groups}
\label{sec::param_ablation}

\begin{figure}[!t]
  \centering
  \includegraphics[width=\linewidth]{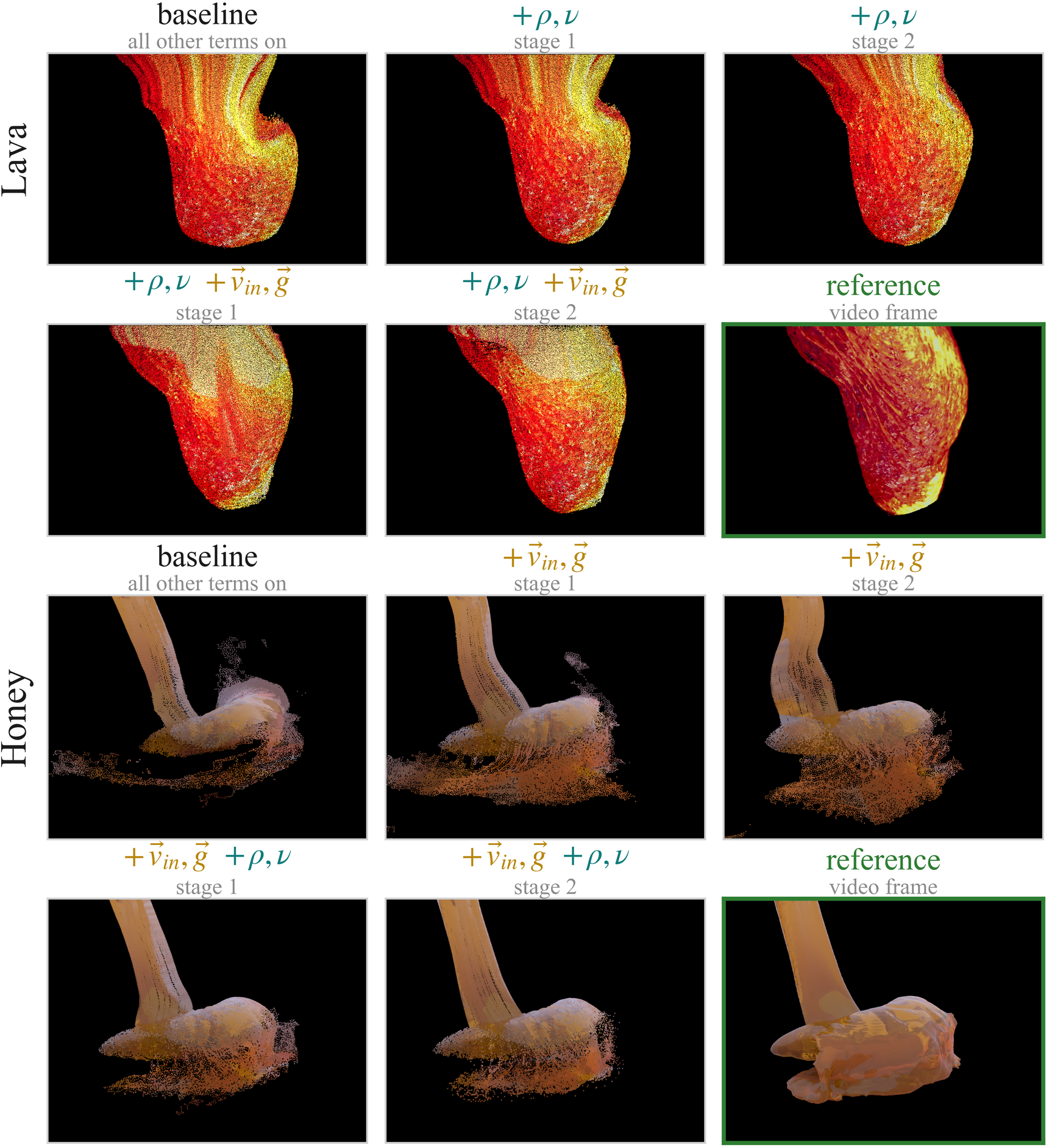}
  \caption{%
    Admitting individual parameter groups to the optimization, on two viscous
    scenes. \emph{Baseline} optimizes all remaining terms of $\Theta$ while
    holding out the annotated group; \emph{stage~1} and \emph{stage~2} continue
    after admitting it.%
  }
  \label{fig:paramablation}
\end{figure}

The terms of $\Theta$ are not interchangeable, and their roles show most clearly
on viscous fluids. In Fig.~\ref{fig:paramablation} each block holds out one
group and then admits it. Without the group the result is still fluid-like
but does not match the input. The material terms $\rho,\nu$ give the bulk its
cohesion and its thickening as the fluid settles, while the driving terms
$\mathbf{v}_{\text{in}},\mathbf{g}$ give the stream its momentum on arrival and
its spread on impact. Only with both groups active does the result
approach the reference. The two blocks admit the groups in opposite orders and end
up alike, so neither is simply absorbing the other's error.

\subsection{User Study} 
\begin{figure}[!t]
    \centering
    \includegraphics[width=\linewidth]{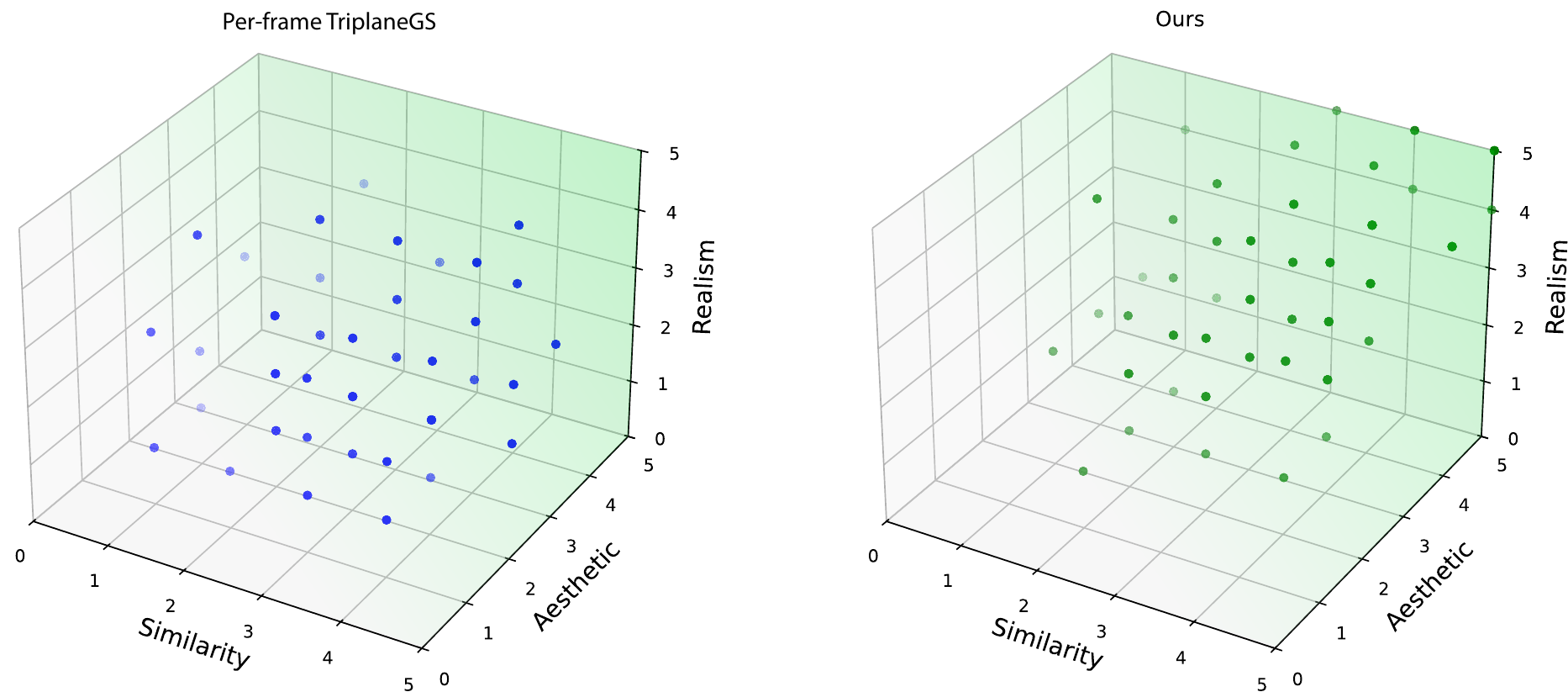}
    \caption{User study scores. Left: continuous per-frame TriplaneGS generation. Right: our dynamic fluid assets. Each point samples one participant-scene pair; greener regions indicate higher scores.}
    \label{fig:userstudy}
\end{figure}

We compare our results against per-frame TriplaneGS~\cite{zou2024triplane} generation, an upper bound on the frame-wise visual similarity that pixel-level fitting (Sec.~\ref{sec:related_deformation}) can reach. Both are sequenced as video and shown side by side, covering the 8 scenes of Fig.~\ref{fig:teaser} and Fig.~\ref{fig:all} (videos in the supplement).
Participants, all full-time 3D artists or graduate students in 3D design, rated each result on a 1-5 scale along three dimensions: (1)~similarity to the input video, (2)~realism, and (3)~aesthetic quality. As shown in Fig.~\ref{fig:userstudy}, the per-frame TriplaneGS baseline achieves slightly higher similarity scores, which is expected since each frame is ad-hoc optimized for pixel-level fidelity. However, our method scores substantially higher in both realism and aesthetics, as the physics-embedded approach produces video-extracted yet natural-looking dynamics rather than pure color space similarity.

\subsection{Evaluation on Different Fluid Types}

We evaluate on both inviscid and viscous liquids.
For inviscid fluids such as rivers and waterfalls (Fig.~\ref{fig:all}~(a)-(e)), gravity and terrain-induced boundary conditions dominate the dynamics. With free-surface splashing, rapid direction changes at terrain features, and high inflow/outflow flux, these cases mainly exercise the boundary coefficients and inlet/outlet velocities in $\Theta$.
For viscous fluids such as honey and lava (Fig.~\ref{fig:all}~(f)--(g)), internal viscous stress constrains relative motion between fluid layers, and boundary damping further attenuates near-wall velocities. The kinematic viscosity $\nu$ and boundary damping $d$ become the dominant parameters. The optimizer finds values that reproduce the characteristically slow, cohesive flow observed in the video.
Additionally, static obstacles such as rocks on the water surface (Fig.~\ref{fig:teaser}, Fig.~\ref{fig:all}~(a),~(e)) are successfully identified and separated during the geometry preprocessing stage, enabling correct collision treatment in the simulation.

\subsection{Compatibility with Different 3DGS Generators}
We demonstrate results with two generators, TriplaneGS~\cite{zou2024triplane} (Fig.~\ref{fig:all}~(a),~(c)-(g)) and TRELLIS~\cite{xiang2024structured} (Fig.~\ref{fig:all}~(b)).
TriplaneGS uses triplane encoding to faithfully preserve input textures, producing Gaussians that closely match the conditioning image at the cost of lower spatial resolution.
TRELLIS, based on rectified flow transformers, generates higher-resolution Gaussians but diverges more from the input image even at high classifier-free guidance, due to its stronger generative prior. On the other hand, it often yields more stereoscopic geometry, particularly beneficial for scenes such as waterfalls that TriplaneGS reconstructs flatter.
Our pipeline processes both into physically plausible fluid assets, benefiting from their respective strengths.

\subsection{User Editing}

\begin{figure}[!t]
    \centering
    \includegraphics[width=\linewidth]{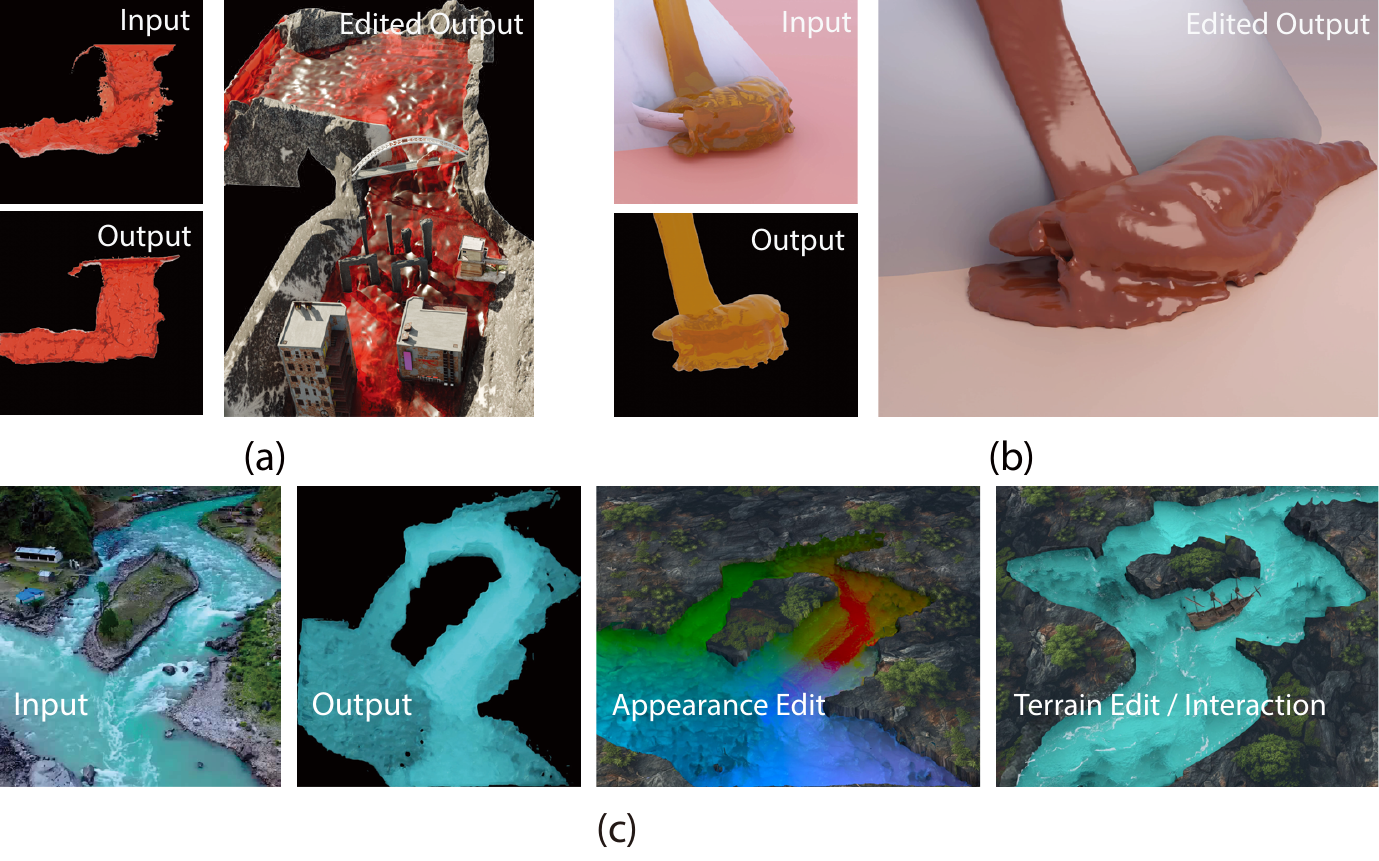}
    \caption{Editing and scene integration. (a)~The extracted river asset is integrated into a digital scene with buildings and a bridge, with extended downstream terrain. (b)~Fluid appearance, viscosity, and gravity are modified for a different visual style. (c)~From left to right: input video, reconstructed output, appearance re-coloring, and terrain editing with an interactive vessel.}%
    \label{fig:editing}
\end{figure}

Governed by explicit physical parameters with finite degrees of freedom, the asset is intuitive to edit: users can adjust fluid material (density, viscosity), body force direction, or inflow/outflow velocities, and the simulation automatically adjusts to produce physically consistent dynamics, as demonstrated in Fig.~\ref{fig:teaser} and Fig.~\ref{fig:editing}.
Since the output also includes the reconstructed terrain, it can be directly integrated into digital scenes for virtual production. Common applications with fluids include placing structures along the flow, introducing interacting objects (e.g. vessels), or reshaping the terrain. The simulation responds accordingly without manual re-animation.

\subsection{Runtime}

\subsubsection{Early Stopping and Physical Convergence}
\label{sec::earlystop}

\begin{figure}[!t]
  \centering
  \includegraphics[width=\linewidth]{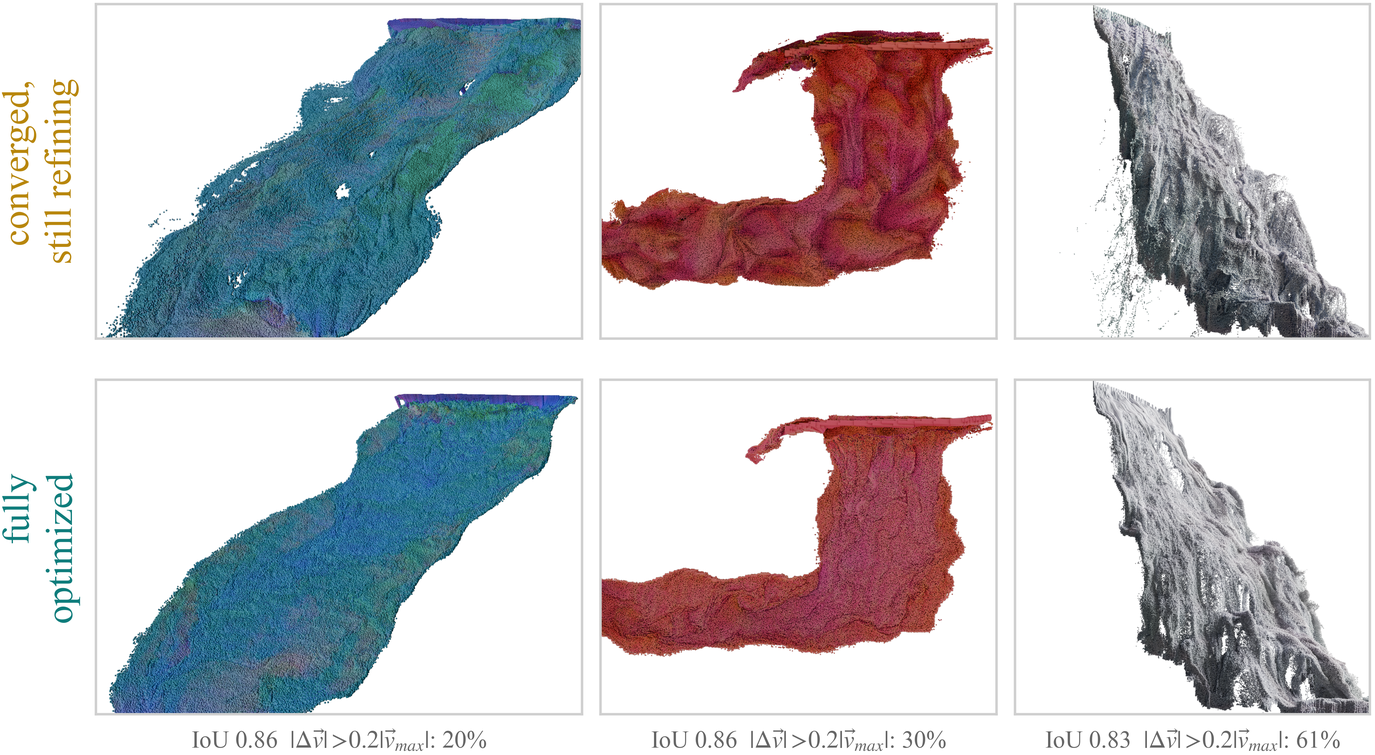}
  \caption{%
    Visual convergence precedes physical convergence. Top: at early stopping.
    Bottom: after the full optimization. Color encodes velocity.%
  }
  \label{fig:earlystop}
\end{figure}

The optimization dominates the runtime. Figure~\ref{fig:earlystop} contrasts
early stopping, where the loss begins to decrease slowly.
Appearance converges well before the dynamics. The two look alike, yet a
substantial fraction of cells still change velocity over the tail, refining flow
rate, direction and local structure.

The cost is therefore a choice, not a fixed price. Where physical fidelity is
not the goal, the early-stopped state is a good place to stop. The flow has
stabilized and reads plausibly, and the run may be resumed from the same basin.
Where the dynamics matter, the tail buys them, and how much remains is not
apparent from appearance. The rightmost scene changes least visually yet most in
velocity. Further reductions remain open: a region-weighted loss, fused solver
kernels, more VRAM for larger batches, and the analytical-gradient variant of
Sec.~\ref{sec::opt_evolution}.

\begin{figure}[!t]
    \centering
    \includegraphics[width=\linewidth]{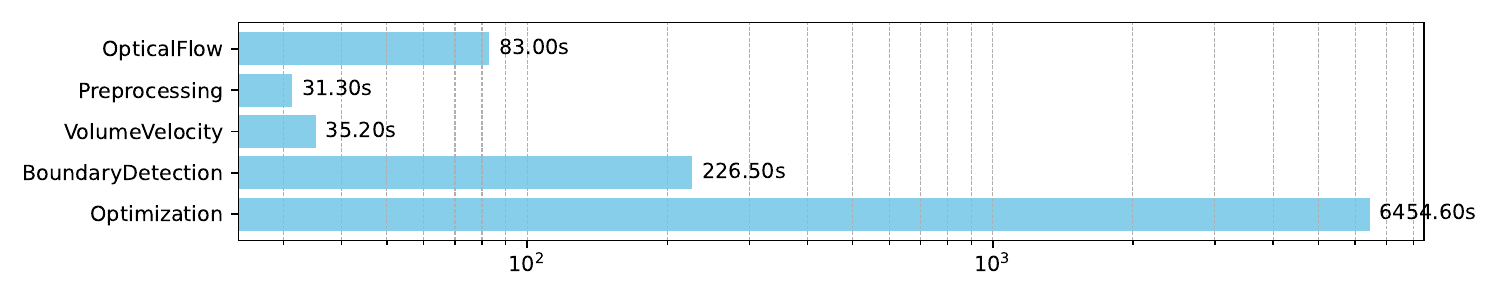}
    \vspace{-10pt}
    \caption{Runtime breakdown (seconds, logarithmic scale) for the scene in Fig.~\ref{fig:teaser}. Parameter optimization dominates the total cost.}
    \label{fig:time}
\end{figure}

Figure~\ref{fig:time} breaks down the runtime of our pipeline for the scene in Fig.~\ref{fig:teaser}.
Preprocessing (geometry denoising, densification, and union) and volumetric velocity construction each complete in under a minute. Optical flow computation and boundary detection (including terrain extraction and wall-function computation) take slightly longer. The differentiable parameter optimization dominates the total runtime, accounting for over 94\% of the cost in this example, because each training sample requires up to 25 forward simulation steps with approximately 80 PeRCNN recurrent layers per step.
The optimization cost scales primarily with two factors: the number of input video frames (which determines the training set size and window length $N$) and the grid resolution (which affects both the simulation step cost and the pressure solve). For scenes with steady-state flow, longer optimization windows can be used without proportionally increasing the cost, as the dynamics converge quickly. For highly dynamic scenes, finer grids and shorter time steps are necessary, increasing the computational burden.
Once the parameters are optimized, the forward APIC re-simulation runs at interactive rates utilizing GPU parallelization.

\section{Conclusions}

We have presented a physics-driven framework for extracting simulation-ready 3D
fluid assets from single-view videos. Across four stages it recovers the
geometry, the surface motion, a volumetric velocity field and a compact set of
explicit simulation parameters, keeping appearance separate from dynamics and 
is not lost to fluid's topological change. A fluid is an open
system with no stable point correspondence, so what can be recovered and
stays useful afterwards is a grid-based field with the parameters that govern
it rather than a per-frame fit. The resulting asset can be re-simulated, edited,
interacted with and extended in time while staying consistent under novel views.
Results on captured and synthetic scenes, different 3DGS generators and a stage-wise
ablation of the extraction chain support the approach.

Our pipeline is staged rather than end-to-end, so each stage rests on the one
before. Thus, it is stage-wise improvable, as a component can
be strengthened without disturbing the others. Several of these dependencies
proved mild in practice. For instance, occlusion in input video can be tolerated wherever the generator places fluid
behind the occluder,
nonideal backside geometry is absorbed by optimizing the boundary conditions except at
intermediate depths, and later the velocity construction responds to previous stages robustly with
plateaus rather than sharp changes (Sec.~\ref{sec::stagewise}). Two are harder
to work around.
Large camera motion is not currently supported, though 3D registration and
ego-motion compensation would address it. Where no motion is measurable at
all, mainstream interpolation has nothing to extend, but a scene with no
visible motion also reduces how wrong the result can look.

Two limits are more fundamental. Splashing and foam involve air entrainment
beyond a single-phase grid, and would need finer resolution together with a
volume-of-fluid treatment of the interface, much as some fluids would need temperature
coupled to density and viscosity. At our resolution the large-eddy structure is
recovered while subgrid detail is not, and refining it is costly as
simulation and pressure solve scale with cell count. Appearance is likewise less reliable than
motion. Reflective and transparent fluids offer little texture to track, and
since the fluid is open and largely interior, new and newly exposed particles
carry no observed color, making a uniform texture the practical choice. Generic
learned estimators did not help, their training data containing almost no fluid,
so a domain-specific model is the component to replace.

Future work will explore adaptive grid refinement and multi-resolution
strategies to capture fine-scale dynamics without proportionally increasing
cost. Longer videos deepen the computation graph and rarely show the scene as a
whole, so they would need segment-wise optimization with 3D alignment.
Extending the framework to gaseous fluids and to learned appearance models for
dynamic surface texture are also promising directions.

\bibliographystyle{IEEEtran}
\bibliography{references}%

\clearpage
\section*{Supplementary Material}
This supplementary material provides detailed derivations referenced in the paper: the 2D divergence constraint used in surface motion estimation (Appendix~\ref{supp:sec:2d-constraint}), and the differentiability proof of the grid-based velocity evolution and analytical gradients of pressure solver, during the parameter optimization (Appendix~\ref{supp:sec:diff-sim}). Appendix~\ref{supp:sec:hyperparams} lists the hyperparameter settings.

\appendices
\section{Derivation of the 2D Divergence Constraint}
\label{supp:sec:2d-constraint}

We derive the relationship between the 2D screen-space divergence of the projected velocity and the out-of-plane depth information, starting from the 3D incompressibility condition. This derivation corresponds to Eq.~\eqref{eq::div2DV}.

\subsection{Projection of 3D Velocity to Screen Space}

Under perspective projection, a 3D point at position  $\mathbf{x}_{3D}$ $(x, y, z)$ in camera space maps to screen coordinate $\mathbf{x}_{2D}$ $(s, t)$:
\[
s = \frac{x}{z}, \qquad t = \frac{y}{z}.
\]
The screen-space velocity $(u_s, u_t)$ is obtained by differentiating the screen coordinates with respect to time. Given the 3D velocity $\mathbf{v} = (v_x,\, v_y,\, v_z)$:
\[
u_s = \frac{ds}{d\tau} = \frac{v_x}{z} - \frac{s\, v_z}{z}, \qquad u_t = \frac{dt}{d\tau} = \frac{v_y}{z} - \frac{t\, v_z}{z},
\]
where $\tau$ denotes time and we have substituted $x = sz$ and $y = tz$.

\subsection{2D Divergence in Screen Coordinates}

The 2D divergence of the screen-space velocity $\mathbf{u}_{2D} = (u_s, u_t)$ is:
\[
\nabla_{2D} \cdot \mathbf{u}_{2D} = \frac{\partial u_s}{\partial s} + \frac{\partial u_t}{\partial t}.
\]
To evaluate these derivatives, we relate screen-space and world-space partial derivatives using the inverse Jacobian of the projection. Since $s = x/z$ and $t = y/z$, at fixed depth we have:
\[
\frac{\partial}{\partial s} = z\,\frac{\partial}{\partial x}, \qquad \frac{\partial}{\partial t} = z\,\frac{\partial}{\partial y}.
\]

\subsection{Expanding the Divergence}

Substituting and expanding the first term:
\[
\frac{\partial u_s}{\partial s} = z\,\frac{\partial}{\partial x}\!\left(\frac{v_x - s\,v_z}{z}\right).
\]
Applying the quotient rule and noting that $\partial s/\partial x = 1/z$, this yields:
\[
\frac{\partial u_s}{\partial s} = \frac{\partial v_x}{\partial x} - \frac{v_z}{z} - \frac{s}{z}\,\frac{\partial v_z}{\partial s}-\frac{u_s}{z}\frac{\partial z}{\partial s}.
\]
An analogous computation for the second term gives:
\[
\frac{\partial u_t}{\partial t} =  \frac{\partial v_y}{\partial y} - \frac{v_z}{z} - \frac{t}{z}\,\frac{\partial v_z}{\partial t} -\frac{u_t}{z}\frac{\partial z}{\partial t}.
\]
Summing both terms:
\begin{equation}
\begin{aligned}
\nabla_{2D} \cdot \mathbf{u}_{2D} =& \frac{\partial v_x}{\partial x} + \frac{\partial v_y}{\partial y} - \frac{2v_z}{z} \\
&- \frac{s}{z}\,\frac{\partial v_z}{\partial s} - \frac{t}{z}\,\frac{\partial v_z}{\partial t} -\frac{u_s}{z}\frac{\partial z}{\partial s} - \frac{u_t}{z}\frac{\partial z}{\partial t} 
.
\end{aligned}
\label{supp:eq:expandeddd}
\end{equation}

\subsection{Applying the Incompressibility Condition}

The 3D divergence-free condition $\nabla \cdot \mathbf{v} = 0$ gives:
\[
\frac{\partial v_x}{\partial x} + \frac{\partial v_y}{\partial y} = -\frac{\partial v_z}{\partial z}.
\]
Substituting into Eq.~\eqref{supp:eq:expandeddd}:
\begin{equation}
\nabla_{2D} \cdot \mathbf{u}_{2D} \!=\! -\frac{\partial v_z}{\partial z} \!-\! \frac{2v_z}{z} \!-\! \frac{s}{z}\frac{\partial v_z}{\partial s} \!-\! \frac{t}{z}\frac{\partial v_z}{\partial t} \!-\! \frac{u_s}{z}\frac{\partial z}{\partial s} \!-\! \frac{u_t}{z}\frac{\partial z}{\partial t}
\label{supp:eq:final}
\end{equation}

Factoring out $1/z$, this can be written compactly as:
\[
\nabla_{2D} \cdot \mathbf{u}_{2D} \!=\! -\frac{1}{z} \! \left( \! s\frac{\partial}{\partial s} \!+\! t\frac{\partial}{\partial t} \!+\! 2 \! \right) \! v_z 
- \frac{1}{z} \! \left( \! u_s\frac{\partial}{\partial s} \!+\! u_t\frac{\partial}{\partial t} \! \right) \! z - \frac{\partial v_z}{\partial z}
\]
The $\partial v_z / \partial z$ term is negligible near a free surface in quasi-steady flow, as gravity is balanced by the hydrostatic pressure gradient ($\partial p/\partial z \!=\! -\rho g$), thus the body force produces no net vertical acceleration. Meanwhile, vertical velocity perturbations originating from interior disturbances (e.g., bottom topography) attenuate with distance from the bed. Since the free surface is subject to zero normal stress from above, no additional forcing drives vertical variation there, and $v_z$ remains approximately uniform near the surface, giving $\partial v_z / \partial z \approx 0$.  This assumption may become inaccurate for highly dynamic surfaces with significant splashing or breaking.

When discretized between consecutive frames, $v_z$ corresponds to the inter-frame depth change $\dot z$, yielding the form stated in Eq.~\eqref{eq::div2DV}:
\[
\nabla_{2D} \cdot \mathbf{u}_{2D} = 
-\frac{1}{z} \langle \mathbf{x}_{2D}, \nabla_{2D}  \dot z \rangle 
- \frac{2}{z}\dot z
-\frac{1}{z} \langle \mathbf{u}_{2D}, \nabla_{2D}  z      \rangle
.
\]
$ z$ and $ \dot z$ are fully determined by the inter-frame depth information from the 3DGS. In regions where depth varies slowly across the screen, the right-hand side approaches zero, and approximate 2D incompressibility is recovered.

\section{Gradient Derivations for Differentiable Simulation}
\label{supp:sec:diff-sim}

This section details the gradient derivation of each step in the grid-based velocity evolution (Sec.~\ref{sec:grid-evolution}). The simulation parameters $\Theta = \{\mathbf{v}_{\text{in}},\tilde{\mathbf{v}}_{\text{in}}, \, $ $ \mathbf{v}_{\text{out}},\, \rho,\, \nu,\, b,\, d,\, \mathbf{g},\, \Delta t\}$ are optimized by backpropagating through the following sequence of operations. We use $\mathbf{v}(\mathbf{x})$ to denote the velocity at grid node $\mathbf{x}$, $v_n$ for the normal component at a boundary, $\vec{n}$ for the outward boundary normal, and $dx$ for the grid spacing.

\subsection{Step 1: Boundary Condition Enforcement}
Boundary conditions are applied to grid cells adjacent to solid boundaries, inlets, and outlets.

 \textit{Solid boundaries:}
If a grid node $\mathbf{x}$ lies at a solid boundary and the velocity is directed into the boundary ($\mathbf{v} \cdot \vec{n} < 0$), the normal component is reflected with a bouncing coefficient $b \in [0,1]$:
\[
v_n(\mathbf{x}) \leftarrow -b \cdot v_n(\mathbf{x})\] \[ \mathbf{v}'(\mathbf{x})=\mathbf{v}(\mathbf{x})-(1+b)\left(\mathbf{v}(\mathbf{x})\cdot\vec{n} \right) \vec{n}
\]
The corresponding gradients are:
\[
\frac{\partial \mathbf{v}(\mathbf{x})}{\partial b} = -v_n(\mathbf{x})\,\vec{n},\]
\[
\frac{\partial \mathbf{v}(\mathbf{x})}{\partial \vec{n}(\mathbf{x})} = -(1+b)\!\left(\vec{n}(\mathbf{x})\mathbf{v}(\mathbf{x})^\mathsf{T} + (\mathbf{v}(\mathbf{x})\cdot\vec{n}(\mathbf{x}))\,\mathbf{I}\right).
\]
Similarly for damping $d$ with tangent directions $\vec{t}_i$:
\[ \mathbf{v}'(\mathbf{x})=\mathbf{v}(\mathbf{x})-d\left(\mathbf{v}(\mathbf{x})\cdot\vec{t}_i \right) \vec{t}_i
\]
The corresponding gradients are:
\[
\frac{\partial \mathbf{v}(\mathbf{x})}{\partial d} = -v_{t_{i}}(\mathbf{x})\,\vec{t}_i,\]
\[
\frac{\partial \mathbf{v}(\mathbf{x})}{\partial \vec{t}_i(\mathbf{x})} = -d\!\left(\vec{t}_i(\mathbf{x})\mathbf{v}(\mathbf{x})^\mathsf{T} + (\mathbf{v}(\mathbf{x})\cdot\vec{t}_i(\mathbf{x}))\,\mathbf{I}\right).
\]

 \textit{Inlet and outlet:}
At inlet and outlet cells, velocities are directly prescribed:
\[
\mathbf{v}(\mathbf{x}) = \mathbf{v}_{\text{in}}(\mathbf{x}), \quad \frac{\partial \mathbf{v}}{\partial \mathbf{v}_{\text{in}}} = \mathbf{I};\qquad
\mathbf{v}(\mathbf{x}) = \mathbf{v}_{\text{out}}(\mathbf{x}), \quad \frac{\partial \mathbf{v}}{\partial \mathbf{v}_{\text{out}}} = \mathbf{I}.
\]

\subsection{Step 2: Body Force Application}

For non-boundary fluid cells, gravity is applied explicitly:
\[
\mathbf{v}(\mathbf{x}) \leftarrow \mathbf{v}(\mathbf{x}) + \mathbf{g}\,\Delta t, \qquad \frac{\partial \mathbf{v}(\mathbf{x})}{\partial \mathbf{g}} = \Delta t\,\mathbf{I}, \qquad \frac{\partial \mathbf{v}(\mathbf{x})}{\partial \Delta t} = \mathbf{g}.
\]

\subsection{Step 3: Velocity Divergence}

The discrete divergence at each fluid cell is computed via central differences:
\[
\text{div}(\mathbf{x}) = \nabla \cdot \mathbf{v}(\mathbf{x}) = \sum_{i=1}^{3} \frac{v_i(\mathbf{x}+\tfrac{dx}{2}\,\mathbf{e}_i) - v_i(\mathbf{x}-\tfrac{dx}{2}\,\mathbf{e}_i)}{dx},
\]
where $\mathbf{e}_i$ denotes the unit vector along dimension $i$. At boundary cells, the divergence is set to zero. The gradient with respect to neighboring velocities is:
\[
\frac{\partial\,\text{div}(\mathbf{x})}{\partial\, v_i(\mathbf{x} \pm \tfrac{dx}{2}\,\mathbf{e}_i)} = \pm\frac{1}{dx}.
\]

\subsection{Step 4: Pressure Solve}

The pressure field enforces incompressibility through the Poisson equation:
\[
\Delta p(\mathbf{x}) = \frac{\rho}{\Delta t}\,\text{div}(\mathbf{x}).
\]
Discretized with a second-order central scheme on a 3D grid, the Jacobi iteration update reads:
\[
p'(\mathbf{x}) = \frac{1}{6}\left(\sum_{i=1}^{6} p(\mathbf{x}_i) - \frac{\rho\,dx^2}{\Delta t}\,\text{div}(\mathbf{x})\right),
\]
where $\mathbf{x}_i$ denotes the six face-adjacent neighbors of $\mathbf{x}$.
The numerical gradients of one iteration step are:
\[
\frac{\partial p'(\mathbf{x})}{\partial p(\mathbf{x}_i)} = \frac{1}{6} \;\;(\text{if } \mathbf{x}_i \text{ is not at B.C.}), \] \[
\frac{\partial p'(\mathbf{x})}{\partial \rho} = -\frac{dx^2}{6\,\Delta t}\,\text{div}(\mathbf{x}), \qquad
\frac{\partial p'(\mathbf{x})}{\partial\,\text{div}(\mathbf{x})} = -\frac{\rho\,dx^2}{6\,\Delta t}.
\]
At boundary cells, a zero-gradient (Neumann) condition is imposed on pressure.

\indent \textit{Analytical gradients:} solve auxiliary system of equations\\
\noindent
Instead of differentiating through the iterative solver, one can derive the gradient by differentiating the continuous Poisson equation directly:
\begin{align}
\nabla^2 p &= \frac{\rho}{\Delta t}\,\nabla \cdot \mathbf{v}, \\[4pt]
\nabla^2\!\left(\frac{\partial p}{\partial \rho}\right) &= \frac{\nabla \cdot \mathbf{v}}{\Delta t}, \\[4pt]
\nabla^2\!\left(\frac{\partial p}{\partial \mathbf{v}}\right) &= \frac{\rho}{\Delta t}\frac{\partial(\nabla \cdot \mathbf{v})}{\partial \mathbf{v}}, \\[4pt]
\nabla^2\!\left(\frac{\partial p}{\partial \Delta t}\right) &= -\frac{\rho\,\nabla \cdot \mathbf{v}}{\Delta t^2}.
\end{align}
Each auxiliary equation has the same Laplacian operator as the original, so the same solver infrastructure can be reused. However, this requires an extra implicit solution per parameter, which is more expensive than the PeRCNN approach adopted in Sec.~\ref{sec:percnn}. 
We also compare their convergence speeds and find analytical gradients can be imprecise at longer simulation steps.

\subsection{Step 5: Pressure Projection and Boundary Re-enforcement}

The velocity is corrected using the pressure gradient to enforce divergence-free flow:
\[
\mathbf{v}(\mathbf{x}) \leftarrow \mathbf{v}(\mathbf{x}) - \frac{\Delta t}{\rho}\,\nabla p(\mathbf{x}).
\]
This applies to cells where both the cell and its neighbor contain fluid. The gradients are:
\[
\frac{\partial \mathbf{v}(\mathbf{x})}{\partial \rho} = \frac{\Delta t\,\nabla p(\mathbf{x})}{\rho^2}, \qquad
\frac{\partial \mathbf{v}(\mathbf{x})}{\partial p(\mathbf{x}_i)} = -\frac{\Delta t}{\rho}\cdot\frac{\pm 1}{dx},
\]
where the sign depends on the direction of the neighbor $\mathbf{x}_i$ relative to $\mathbf{x}$.

After the pressure projection, the boundary condition from Step~1 is re-applied to cells adjacent to solid boundaries, using the same reflection and damping formulas and gradients.

\section{Hyperparameters}
\label{supp:sec:hyperparams}

\begin{table*}[!t]
\caption{Hyperparameter settings. \textit{F}: fixed algorithm constant.
\textit{I}: initial value of a parameter in $\Theta$, subsequently optimized.
\textit{S}: scene-dependent, representative value shown.}
\label{supp:table:hyper}
\centering
\small
\setlength{\tabcolsep}{4pt}
\begin{tabular}{@{}llcc@{}}
\toprule
\textbf{Stage} & \textbf{Parameter} & \textbf{Value} & \\
\midrule
\textit{Geometry (Sec.~\ref{sec:geometry})} & & & \\
  & Opacity threshold for denoising          & $0.02$      & F \\
  & Downward filling depth $\max_{p_y}$      & $30$ cells  & F \\
  & Cross-frame union: particles per cell    & $1$         & F \\
  & Cross-frame union: frames                & $8$         & F \\
  & Density field grid (cubic)               & $320^3$    & F \\
\addlinespace[3pt]
\textit{Surface motion (Sec.~\ref{sec:surface-motion})} & & & \\
  & Frame step  & $\Delta t_{\text{video}} = 2/\text{fps}$ & F \\
  & Farneb\"ack pyramid scale / levels       & $0.5$ / $3$ & F \\
  & Farneb\"ack window size                  & $15$\,px    & F \\
  & Farneb\"ack polynomial $n$ / $\sigma$    & $5$ / $1.1$ & F \\
  & Farneb\"ack iterations                   & $3$         & F \\
  & Mainstream interp.: detection threshold  & $1.0$\,px   & F \\
  & Mainstream interp.: direction weight power & $3$       & F \\
  & Mainstream interp.: distance weight      & $0.4$       & F \\
  & Mainstream interp.: neighborhood radius  & $0.1\,\max(W,H)$ & F \\
  & 2D constraint: propagation radius        & $0.1\,N_{\max}$ & F \\
  & 2D constraint: propagation decay         & $5$         & F \\
  & Singular velocity clamp                  & $2\Delta x/\Delta t$ & F \\
\addlinespace[3pt]
\textit{Volumetric velocity (Sec.~\ref{sec:volumetric})} & & & \\
  & Boundary layer thickness $\delta$        & $6$ cells   & S \\
  & Near-wall damping factor                 & $0.1$       & F \\
  & Div.-free proj.: strength                & $1.0$       & F \\
  & Div.-free proj.: iteration number        & $100$       & F \\
\addlinespace[3pt]
\textit{Parameter optimization (Sec.~\ref{sec:optimization})} & & & \\
  & Simulation grid                          & $126{\times}37{\times}128$ & S \\
  & Loss weight: direction $\alpha$          & $0.67$      & F \\
  & Loss weight: $L_2$ $\beta$                & $0.33$      & F \\
  & Optimizer                                & Adam        & F \\
  & Learning rate                            & $10^{-3}$   & F \\
  & Learning-rate schedule                   & $\times 0.95$ / $100$ ep. & F \\
  & Gradient magnitude clip                  & $0.35$      & F \\
  & Training epochs                          & $200$       & F \\
  & Training samples per batch               & $5$         & F \\
  & Frame offset per sample                  & $12$        & F \\
  & Simulation steps per training sample     & $25$        & F \\
  & PeRCNN recurrent layers per step         & $\approx 80$ & F \\
  & Convergence criterion                    & loss $<5\times10^{-6}$ & F \\
  & Density $\rho$                           & $10^{3}$    & I \\
  & Time step ratio $\Delta t$               & $1.0$       & I \\
  & Div.-free proj.: terrain restitution $b$ & $0.5$       & I \\
\bottomrule
\end{tabular}
\end{table*}

Table~\ref{supp:table:hyper} lists the settings of every stage. We distinguish three
kinds of entries. \textbf{Fixed} values are algorithm constants shared by all
scenes. \textbf{Initial} values are members of the optimized parameter set
$\Theta$ and are only the starting point of the optimization, whose converged
values differ per scene (Sec.~\ref{sec:optimization}). \textbf{Scene} values
follow the captured geometry and are reported for the river of Fig.~\ref{fig:overview} as a
representative case. The sensitivity of the most influential parameters is
reported in Fig.~\ref{fig:stagesensitivity}. The loss weights favor direction over
magnitude, direction being the quantity invariant to the scale ambiguity of a
monocular reconstruction.

Two trainings take place, on different supervision signals. The PeRCNN kernels
are fitted against pressure, and the physical parameters $\Theta$ are then
optimized against velocity.

Fitting the PeRCNN kernels (Sec.~\ref{sec:percnn}) precedes the
parameter optimization and is not repeated during it, so its cost does not enter
the runtime breakdown. A fitted set also transfers to scenes of similar geometry and grid. The pressure fields recorded before and after a standard
implicit solve supply the input and the target, over $128$ frames of the
reconstructed field, under an MSE loss. It uses Adam at a learning rate of
$10^{-2}$, decayed by $0.91$ every $100$ epochs, for $5000$ epochs. The pressure
and boundary operators are $3\times3\times3$ convolutions masked to the
finite-difference stencil, and the divergence operator is a $2\times2\times2$
convolution on the MAC grid. The fitted weights converge as reported in
Fig.~\ref{fig:percnn_str_fit}.

The parameter optimization then fits $\Theta$ to the reconstructed velocity
field under Eq.~\eqref{eq::optLoss} with the kernels held fixed, using the
\textit{Parameter optimization} rows of Table~\ref{supp:table:hyper}.

\end{document}